\documentclass[%
reprint,
superscriptaddress,
amsmath,amssymb,
aps,
prb,
]{revtex4-2}

\usepackage{graphicx}
\usepackage{dcolumn}
\usepackage{bm}
\usepackage{amsmath, amssymb}
\usepackage{makecell}
\usepackage{multirow}
\usepackage{float}
\usepackage{color}

\usepackage{tikz}
\usetikzlibrary{shapes.geometric}

\definecolor{markerB}{rgb}{0.35,0.65,1.00}
\definecolor{markerH}{rgb}{0.20,0.50,0.88}
\definecolor{markerD}{rgb}{0.08,0.32,0.70}

\definecolor{markerA}{rgb}{1.00,0.00,0.00}
\definecolor{markerG}{rgb}{0.68,0.16,0.20}

\definecolor{markerE}{rgb}{0.38,0.56,0.15}
\definecolor{markerF}{rgb}{0.28,0.44,0.10}

\definecolor{markerC+}{rgb}{0.86,0.50,0.50}
\definecolor{markerCc}{rgb}{0.60,0.78,0.95}

\newcommand{\BICasterisk}{%
	\tikz[baseline=-0.55ex]{
		\draw[markerA,line width=0.8pt]
		(0:1.3mm)--(180:1.3mm)
		(45:1.3mm)--(225:1.3mm)
		(90:1.3mm)--(270:1.3mm)
		(135:1.3mm)--(315:1.3mm);
	}
}

\newcommand{\BICstarPoly}[2]{%
	\tikz[baseline=-0.55ex]{
		\node[
		star,
		star points=#1,
		star point ratio=2.25,
		draw=#2,
		fill=#2,
		minimum size=3.5mm,
		inner sep=0pt
		] {};
	}
}

\newcommand{\BICpoly}[3]{%
	\tikz[baseline=-0.55ex]{
		\node[
		regular polygon,
		regular polygon sides=#1,
		rotate=#3,
		draw=#2,
		fill=#2,
		minimum size=3.1mm,
		inner sep=0pt
		] {};
	}
}

\newcommand{\BICcross}[1]{%
	\tikz[baseline=-0.55ex]{
		\draw[#1,line width=0.8pt]
		(-1.2mm,-1.2mm)--(1.2mm,1.2mm)
		(-1.2mm,1.2mm)--(1.2mm,-1.2mm);
	}
}

\newcommand{\BICdouble}{%
	\tikz[baseline=-0.55ex]{
		\draw[markerC+,line width=0.8pt]
		(0,0) circle (1.35mm);
		\draw[markerCc,line width=0.8pt]
		(-1.2mm,0)--(1.2mm,0)
		(0,-1.2mm)--(0,1.2mm);
	}
}

\makeatletter

\newcommand{\Rmnum}[1]{\expandafter\@slowromancap\romannumeral #1@}

\makeatother                                                                   

\begin{document}                                                                
\preprint{APS/123-QED}

\title{Super bound states in the continuum: Classification, direct computation, and geometry in parameter space} 

\author{Nan Zhang}
\email{nan25@vt.edu}
\affiliation{Bradley Department of Electrical and Computer Engineering, Virginia Tech, Blacksburg, Virginia 24061, USA
}
\affiliation{Department of Mathematics, City University of Hong Kong, Kowloon, Hong Kong, China
}
\author{Ya Yan Lu}%
\affiliation{Department of Mathematics, City University of Hong Kong, Kowloon, Hong Kong, China
}

\date{\today}

\begin{abstract}
In periodic structures, resonant states near a bound state in the
continuum (BIC) have quality ($Q$) factors scaling as
$Q\sim 1/\delta^{2\mu}$, where $\delta$ is the wavevector
detuning from the BIC.
While generic BICs have $\mu=1$,
super-BICs with $\mu\geq 2$ give rise to higher-$Q$
resonances over a broad wavevector range,
making them attractive for enhancing light-matter interactions.
Here, we develop a systematic theoretical 
and computational framework for studying super-BICs.
We first classify them according to $\mu$ and symmetry.
We then study their unfolding under structural variations,
characterizing the bifurcations of nearby generic BICs
and determining the minimum number $n$ of additional tunable structural parameters
required to recover a super-BIC of the same class.
We devise a direct computational method for locating super-BICs together with their associated structural parameters and efficiently tracing their families.
Finally, we investigate the local geometry of the BIC set in the product space of structural parameters and the Brillouin zone.
This geometric framework provides a unified interpretation of the unfolding of super-BICs and offers a new understanding of BIC merging and splitting phenomena.
Our results establish a theoretical and computational foundation
for the systematic study of super-BICs,
with potential applications in resonant optics
and photonic device design.
\end{abstract}
\maketitle

\section{Introduction}

Bound states in the continuum (BICs) are trapped modes existing within the continuous spectrum of radiation states~\cite{JvonNeumann1929PZ,HFriedrich1985PRA,ABonnetBendhia1994MMAS,DEvans1994JFM,CHsu2016NRM,KKoshelev2023PU}.
This counterintuitive phenomenon was first theoretically explored by
von Neumann and Wigner in a quantum system~\cite{JvonNeumann1929PZ}. 
In recent years, BICs have attracted much attention in photonics due to their remarkable properties and extensive applications~\cite{CHsu2016NRM,KKoshelev2023PU,DMarinica2008PRL,EBulgakov2008PRB,YPlotnik2011PRL,CHsu2013Nature,MKang2023NRP}. 
Experimental observations of BICs have been reported across a variety of optical systems, including periodic structures~\cite{CHsu2013Nature}, waveguides with lateral leakage channels~\cite{MWebster2007PTL,CZou2015LPR,ZYu2019Optica}, and anisotropic layered structures~\cite{JGomisBresco2017NP}. Notably, a BIC is often surrounded by a family of resonant states with quality ($Q$) factors approaching infinity~\cite{CHsu2013Nature,JLee2012PRL}. 
These high-$Q$ resonances, along with the associated local field enhancement, 
can be utilized to boost light-matter interactions and facilitate the design of novel photonic devices~\cite{VMocella2015PRB,AKodigala2017Nature,FYesilkoy2019NP,SAzzam2021AOM}.

In a periodic structure, such as a photonic crystal (PhC) slab,
the asymptotic behavior of the $Q$ factors of resonant states near a generic BIC follows an inverse-square law,
i.e., $Q\sim 1/\delta^2$, where $\delta$ denotes a small wavevector detuning from the BIC~\cite{LYuan2017PRA}.
Perturbation theory, which expands the resonant states in powers of $\delta$ about the BIC, 
shows that this scaling is associated with nonzero far-field radiation of the first-order field correction~\cite{LYuan2017PRA,LYuan2020PRAPerturbation,NZhang2025PRL}.
For certain special BICs, the first-order radiation vanishes, 
and the leading-order radiation appears in the $\mu$-th field correction with $\mu\geq 2$, leading to
$Q\sim 1/\delta^{2\mu}$~\cite{LYuan2017PRA,LYuan2020PRAPerturbation,NZhang2025PRL}.
These special BICs are referred to as super-BICs by some researchers~\cite{MHwang2021NC,EBulgakov2023PRB,NLe2024PRL,NZhang2025PRL}.
Compared to generic BICs, the $Q$ factors of resonant states near a super-BIC are further enhanced over a broad wavevector range~\cite{LYuan2017PRA,LYuan2020PRAPerturbation,NZhang2025PRL,MHwang2021NC,EBulgakov2023PRB,NLe2024PRL,EBulgakov2017PRA,JJin2019Nature,MKang2021PRL}.
Recent experimental studies also suggest that super-BICs give rise to ultrahigh-$Q$ resonances that are robust against finite-size effects and fabrication imperfections~\cite{JJin2019Nature,MHwang2021NC,NLe2024PRL,ZZheng2024IEEEPJ}.
These important properties have been exploited to enhance nonlinear effects~\cite{LYuan2017PRA} and design ultrahigh-$Q$ cavities, as well as ultralow-threshold and ultracompact lasers~\cite{AKodigala2017Nature,JJin2019Nature,MHwang2021NC,ZZheng2024IEEEPJ,JCui2025NP}.

Super-BICs are generally non-generic and require parameter tuning for their realization~\cite{EBulgakov2017PRA,JJin2019Nature,MKang2021PRL,LYuan2017PRA,LYuan2020PRAPerturbation,NZhang2025PRL,NZhang2024OE,NZhang2023OE}.
A common approach is to merge multiple BICs on a single dispersion surface by tuning structural parameters~\cite{EBulgakov2017PRA,JJin2019Nature,MKang2021PRL,EBulgakov2023PRB,MKang2022LSA}.
At the coalescence point, the $Q$-factor scaling changes to a higher-order inverse power law, yielding a super-BIC~\cite{JJin2019Nature}.
For this reason, such super-BICs are often referred to as merging-BICs~\cite{JJin2019Nature,MKang2021PRL,MKang2022LSA}.
However, it has also been shown that the same super-BIC may exhibit BIC merging under one parameter variation but not under another~\cite{NZhang2024OL}.

Existing studies of super-BICs in periodic
structures~\cite{LYuan2017PRA,JJin2019Nature,LYuan2020PRAPerturbation,MKang2021PRL,MHwang2021NC,
	MKang2022LSA,EBulgakov2023PRB,HLuo2023PRA,SLee2023LPR,NShubin2023PRB,
	HBarkaoui2023PRB,NZhang2024OL,NLe2024PRL,LLiu2024OE,ZZheng2024IEEEPJ,
	NZhang2025PRL,JCui2025NP}
have presented a rich variety of examples
across different structural configurations.
In this work, we develop a systematic theoretical
and computational framework for studying super-BICs.
We first classify super-BICs according to the
leading radiation order $\mu$ of nearby resonant
states and the symmetries of the structure.
Since super-BICs are generally not preserved
under structural perturbations,
we next study their unfolding.
We develop a general bifurcation theory to characterize
how nearby generic BICs emerge
under structural variations.
We then establish a theory of parametric dependence
to determine the minimum number $n$ of additional
tunable structural parameters required to recover
a nearby super-BIC of the same class.
This number also gives the minimum number of
tunable structural parameters required to realize that class.
Previous studies of the parametric dependence of
generic BICs and bifurcations of certain super-BIC
classes provide a starting point~\cite{LYuan2020PRAParametric,NZhang2024OE,NZhang2024OL}.

Guided by the classification and unfolding analysis,
we devise a direct computational method for locating
super-BICs together with their associated structural parameters
and tracing their families in structural parameter space.
By introducing an auxiliary eigenvalue problem,
the method identifies super-BICs through
the simultaneous vanishing of the auxiliary eigenvalue
and lower-order radiation coefficients,
without tracking the merging of multiple BICs.
This approach is particularly advantageous for super-BICs
with higher values of $\mu$, whose realization
through BIC merging can be challenging~\cite{JJin2019Nature,HLuo2023PRA,LLiu2024OE}.

Finally, we investigate the local geometry of the BIC set
near super-BICs in the product space of structural
parameters and the Brillouin zone.
This geometric framework provides a unified interpretation
of the unfolding of super-BICs.
The number $n$ is interpreted as the codimension
of the corresponding super-BIC submanifold
within the BIC manifold, while different structural
paths select different sections of the BIC set
and determine the behavior of nearby generic BICs.
In particular, non-generic paths may lead to
qualitatively different bifurcation behaviors.
This geometric viewpoint also provides a new understanding
of BIC merging and splitting phenomena.

The systematic framework
developed in this paper applies to structures
with various in-plane symmetries
and covers both nondegenerate and degenerate BICs,
at-$\Gamma$ and off-$\Gamma$ configurations,
and arbitrary values of $\mu$.
These results may inform the design of photonic structures
supporting ultrahigh-$Q$ resonances for enhanced light-matter interactions.

\begin{figure}[t]
	\centering
	\includegraphics[scale=0.31]{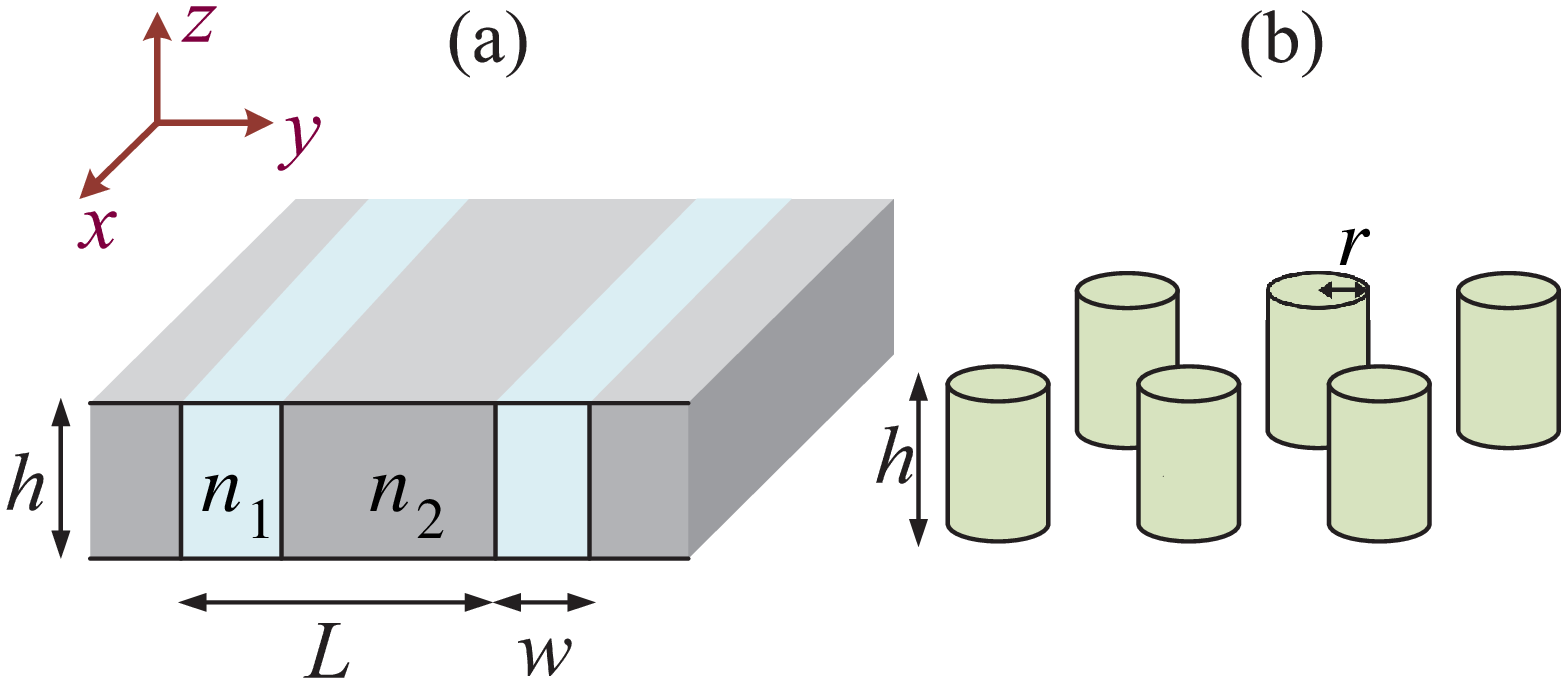}
	\caption{Schematic illustrations of (a) a diffraction grating and (b) a square lattice of dielectric rods, 
		both embedded in a homogeneous medium. 
		The grating is a 2D periodic structure that is invariant in $x$ and periodic in $y$ with period $L$. 
		It consists of two dielectric materials of the same height $h$, 
		with refractive indices $n_1$ and $n_2$ and widths $w$ and $L-w$, respectively. 
		The square lattice is a 3D biperiodic structure with period $L$ in both $x$ and $y$, 
		where the rods have height $h$ and radius $r$.}\label{structure}
\end{figure}

To simplify the presentation,
we first develop our theory in the scalar setting, 
specifically $E$-polarized super-BICs in two-dimensional (2D) lossless periodic structures 
that are invariant along one direction and have one-dimensional (1D) periodicity [e.g., Fig.~\ref{structure}(a)].
We then extend the results to the vectorial setting,
i.e., super-BICs in three-dimensional (3D) biperiodic structures [e.g., Fig.~\ref{structure}(b)].

The paper is organized as follows.
In Secs.~\Rmnum{2}--\Rmnum{6}, we focus on the scalar setting.
In Secs.~\Rmnum{2} and \Rmnum{3}, we present the basic theory and classification of super-BICs.
In Sec.~\Rmnum{4}, we study the unfolding of super-BICs
by developing the bifurcation theory and the theory of parametric dependence.
In Sec.~\Rmnum{5}, we present the direct computational method
and validate it with numerical examples.
In Sec.~\Rmnum{6}, we investigate the local geometry of the BIC set near super-BICs.
In Sec.~\Rmnum{7}, we extend the main results to vectorial fields.
Finally, Sec.~\Rmnum{8} concludes the paper.

\section{Super-BICs in 2D structures with 1D periodicity}

In this section, we review the relation between the leading radiation of
resonant states near an $E$-polarized BIC and the asymptotic scaling of their
$Q$ factors in a 2D structure with 1D periodicity
~\cite{NZhang2025PRL,NZhang2026arXiv}.
This relation provides the basis for the classification of super-BICs developed in
Sec.~\Rmnum{3}.

We consider a 2D lossless, isotropic, and nonmagnetic structure that is invariant in $x$,
periodic in $y$ with period $L$, bounded in $z$,
and embedded in a homogeneous background medium with dielectric constant
$\varepsilon_b$.
The structure is assumed to have reflection symmetries in both $y$ and $z$.
We use Heaviside--Lorentz units with $c=1$ and the time dependence
$\exp(-i\omega t)$. An $E$-polarized electromagnetic field has the form
\begin{equation}
	{\bm E}
	=
	\begin{bmatrix}
		E_x\\
		0\\
		0
	\end{bmatrix},
	\quad
	{\bm H}
	=
	\begin{bmatrix}
		0\\
		H_y\\
		H_z
	\end{bmatrix}.
\end{equation}
Denote $u({\bm r}):=E_x({\bm r})$, where ${\bm r}=(y,z)$.
The field $u$ satisfies the scalar Helmholtz equation
\begin{equation}\label{mainEq}
	{\cal L}u
	:=
	\left[
	\partial_y^2+\partial_z^2
	+\omega^2\varepsilon({\bm r})
	\right]u
	=
	0,
\end{equation}
where $\varepsilon({\bm r})$ is the dielectric function.
For a Bloch mode with wavenumber
$-\pi/L\leq\beta<\pi/L$, $u$ satisfies
\begin{equation}
	u(y+L,z)=e^{i\beta L}u(y,z).
\end{equation}

A resonant state has a real Bloch wavenumber $\beta$
and a complex frequency $\omega$ with $\operatorname{Im}(\omega)<0$,
and satisfies the outgoing radiation condition.
The $Q$ factor is defined by
$Q=-0.5\operatorname{Re}(\omega)/\operatorname{Im}(\omega)$.
We assume that only the zeroth diffraction order is propagating in the
background medium.
Let ${\bm k}^{\pm}=(\beta,\pm\gamma)$, where
$\gamma=\sqrt{\omega^2\varepsilon_b-\beta^2}$.
The outgoing far-field radiation patterns are given by
\begin{equation}\label{FFPatternU}
	u({\bm r})
	\sim
	d^\pm e^{i{\bm k}^{\pm}\cdot{\bm r}},
	\quad
	z\rightarrow\pm\infty.
\end{equation}
The up-down reflection symmetry implies either $d^+=d^-$ or $d^+=-d^-$.
As $\beta$ varies, resonant states above the light line form bands.
A BIC is an isolated point on such a band at $\beta=\beta_*$,
for which the corresponding field $u_*({\bm r})$ is localized with
vanishing radiation coefficients, $d^+=d^-=0$.
Since there is no radiation loss, the frequency $\omega_*$ is real and the
$Q$ factor is infinite.

\subsection{Nondegenerate case}

We next study the relation between the leading radiation of resonant states
near a BIC and the asymptotic scaling of their $Q$ factors.
We first consider a nondegenerate BIC.
Let $\delta=(\beta-\beta_*)L$ denote the dimensionless detuning from the BIC.
For sufficiently small $|\delta|$, the frequency and field can be expanded as
\begin{align}
	\label{ExpansionFre}
	&\omega
	=
	\omega_*
	+
	\delta\omega_1
	+
	\delta^2\omega_2
	+\cdots,
	\\
	\label{ExpansionU}
	&u
	=
	e^{i\delta y/L}
	\left(
	u_*
	+
	\delta u_1
	+\cdots
	\right),
\end{align}
where all $u_j$ have the same quasi-periodicity as $u_*$.

Substituting Eqs.~\eqref{ExpansionFre} and \eqref{ExpansionU} into
Eq.~\eqref{mainEq} and collecting terms of order ${\cal O}(\delta^j)$
yield a sequence of perturbation equations
\begin{equation}
	\label{EqLj}
	{\cal L}_*u_j
	=
	f_j\left({\bm r};\omega_j\right),
\end{equation}
where ${\cal L}_*$ denotes the operator ${\cal L}$ evaluated at $\omega=\omega_*$, 
and $f_j$ depends on the lower-order solutions 
and is linear in the frequency correction $\omega_j$.
We normalize the BIC by
\begin{equation}
	\left\langle u_*|\varepsilon|u_*\right\rangle
	:=
	\frac{1}{L^2}
	\int_\Omega
	\overline{u}_*({\bm r})
	\,\varepsilon({\bm r})\,
	u_*({\bm r})
	\,{\rm d}{\bm r}
	=
	1,
\end{equation}
where $\Omega=(-L/2,L/2)\times\mathbb{R}$ is a unit cell.
At each order, the solvability condition
$\langle u_*|f_j\rangle=0$
determines $\omega_j$.
The field correction $u_j$ is then obtained uniquely from
Eq.~\eqref{EqLj} by imposing the gauge condition
$\langle u_*|\varepsilon|u_j\rangle=0$.

We examine the far-field radiation patterns of the perturbation solutions.
Due to the up-down reflection symmetry, it is sufficient to consider
the upward radiation, and we henceforth omit the superscript ``$+$''.
The radiation coefficient $d$ in Eq.~\eqref{FFPatternU} can be
expanded in powers of $\delta$ as
\begin{equation}
	d
	=
	\delta d_1
	+\delta^2 d_2
	+\delta^3 d_3
	+\cdots .
	\label{eq:d_expansion}
\end{equation}
The vertical wavenumber $\gamma$ can likewise be expanded in powers of
$\delta$.
Let ${\bm k}_*=(\beta_*,\gamma_*)$, where
$\gamma_*=\sqrt{\omega_*^2\varepsilon_b-\beta_*^2}>0$.
Substituting the expansions of $u$, $d$, and $\gamma$ into
the far-field expression~\eqref{FFPatternU} and matching powers of $\delta$ yield
\begin{equation}
	\label{FFPatternUJ}
	u_j({\bm r})
	\sim
	\hat d_j(z)e^{i{\bm k}_*\cdot{\bm r}},
	\quad
	z\rightarrow\infty,
\end{equation}
where $\hat d_j(z)$ is a polynomial in $z$ of degree at most $j-1$,
whose constant term is $d_j$.
Suppose that $u_\mu({\bm r})$, with $\mu\geq1$,
is the leading perturbation correction that radiates
to infinity.
Then $\hat d_j(z)=0$ for $1\leq j<\mu$ and
$\hat d_\mu(z)=d_\mu\neq0$.
Consequently, the outgoing radiation coefficients satisfy
\begin{equation}
	d_1=\cdots=d_{\mu-1}=0,
	\quad
	d_\mu\neq0.
\end{equation}

We relate $d_\mu$ to the asymptotic scaling of the $Q$ factors of nearby
resonant states.
Multiplying Eq.~\eqref{mainEq} by $\overline{u}$,
integrating over one periodic cell truncated in the $z$ direction,
and applying Green's identity yield an energy-conservation relation
between the decay rate, the stored energy, and the outgoing radiation power.
We expand this relation in powers of $\delta$ at fixed truncation
and then take the truncation boundaries to infinity order by order.
Since the first $\mu-1$ field corrections have no outgoing radiation,
we obtain
\begin{equation}
	\operatorname{Im}(\omega_1)
	=
	\cdots
	=
	\operatorname{Im}(\omega_{2\mu-1})
	=
	0.
\end{equation}
At leading order, energy conservation gives
\begin{equation}
	-2U_*\operatorname{Im}(\omega_{2\mu})
	=
	P_{2\mu},
	\quad
	P_{2\mu}
	=
	\frac{\gamma_*L}{\omega_*}|d_\mu|^2,
\end{equation}
where $U_*=L^2/2$ is the stored energy of the BIC
under our normalization, and $P_{2\mu}$ is the leading
radiation power.
Consequently,
\begin{equation}\label{Q2nu}
	Q
	\sim
	\frac{Q_{2\mu}}{\delta^{2\mu}},
	\quad
	Q_{2\mu}
	=
	\frac{\omega_* U_*}{P_{2\mu}}.
\end{equation}
Thus, the leading radiation order $\mu$ determines the
$Q$-factor scaling exponent, while $d_\mu$ determines
the coefficient $Q_{2\mu}$.
For a generic BIC, $\mu=1$, whereas a super-BIC
is characterized by $\mu\geq2$.

\subsection{Degenerate case}

Next, we extend the above analysis to degenerate BICs.
Consider an $m$-fold degenerate BIC with a set of orthonormal states
$\phi_1,\ldots,\phi_m$.
The BIC associated with a nearby resonant band 
is a linear combination of the degenerate states $\phi_1,\ldots,\phi_m$, 
with coefficients collected in the vector ${\bm c}=[c_1,\ldots,c_m]^{\sf T}$.
The normalization of $u_*$ gives
${\bm c}^\dagger{\bm c}=1$.
The expansions in Eqs.~\eqref{ExpansionFre} and \eqref{ExpansionU}
and the perturbation equations~\eqref{EqLj}
remain valid.
At first order, the solvability conditions
$\langle \phi_\ell|f_1\rangle=0$ for $1\leq\ell\leq m$
lead to the eigenvalue problem
\begin{equation}\label{degenerateomega1}
	\left(
	{\bm A}-\omega_1{\bm I}_m
	\right)
	{\bm c}
	=
	0,
\end{equation}
where ${\bm I}_m$ is the $m\times m$ identity matrix and
${\bm A}$ is the Hermitian matrix with elements
\begin{equation}\label{matrixA}
	A_{\ell s}
	=
	-\frac{i}{\omega_*L}
	\left\langle
	\phi_\ell\middle|\partial_y\middle|\phi_s
	\right\rangle .
\end{equation}
If ${\bm A}\neq0$, each eigenpair $(\omega_1,{\bm c})$ of
Eq.~\eqref{degenerateomega1} generally determines a resonant band
emanating from the degenerate BIC.

We consider a simple eigenpair $(\omega_1,{\bm c})$.
Let $v_1$ be a particular solution of the first-order perturbation equation.
The general first-order field correction $u_1$ 
is the sum of a particular solution $v_1$ 
and a linear combination of the degenerate states $\phi_1,\ldots,\phi_m$, 
with coefficients collected in the vector ${\bm c}_1=[c_{11},\ldots,c_{1m}]^{\sf T}$.
The solvability conditions for the second-order perturbation equation give
\begin{equation}\label{equationu1c1}
	\left(
	{\bm A}-\omega_1{\bm I}_m
	\right)
	{\bm c}_1
	=
	\omega_2{\bm c}
	-
	{\bm b}_1,
\end{equation}
where ${\bm b}_1$ is determined by the lower-order perturbation solutions.
The solvability condition for the linear system~\eqref{equationu1c1}
then gives
$\omega_2={\bm c}^\dagger{\bm b}_1$.
Equation~\eqref{equationu1c1}, together with the gauge condition $\langle u_*|\varepsilon|u_1\rangle=0$,
uniquely determines ${\bm c}_1$ and hence $u_1$.

The same procedure can be continued order by order.
In the special case ${\bm A}=0$, the first-order problem gives
$\omega_1=0$ but does not determine ${\bm c}$, which is then determined
by the solvability conditions at higher orders.
In either case, the far-field radiation patterns are obtained as in the
nondegenerate case, and the leading radiation order $\mu$ and its
corresponding coefficient $d_\mu$ determine the asymptotic
$Q$-factor scaling of each resonant band.

\section{Classification of super-BICs}

In this section, we classify super-BICs in 2D structures with 1D
periodicity.
A BIC can be located either at $\Gamma$, with $\beta_*=0$, or off $\Gamma$,
with $\beta_*\neq0$.
For a nondegenerate at-$\Gamma$ BIC, the reflection symmetry in $y$
allows the BIC to be chosen as either a symmetric standing wave (SSW),
satisfying
$u_*(y,z)=u_*(-y,z)$, or an antisymmetric standing wave (ASW),
satisfying
$u_*(y,z)=-u_*(-y,z)$.

For an SSW, $f_j$ in Eq.~\eqref{EqLj} is odd in $y$ for odd $j$
and even in $y$ for even $j$.
Accordingly, $u_j$ has the same parity as $f_j$.
In particular, $u_1$ is odd in $y$ and therefore has no outgoing
zeroth-order radiation, so that $d_1=0$.
Thus, an SSW is automatically a super-BIC.
More generally, the radiation coefficients vanish at all odd orders,
and the leading radiation order satisfies
$\mu=2,4,6,\ldots$.
The case $\mu=2$ was first reported and rigorously justified in
Ref.~\cite{LYuan2017PRA}, where an SSW was also used to achieve optical
bistability at low incident intensity.
To our knowledge, the case $\mu=4$ has not previously been reported in
2D structures.
In Sec.~\Rmnum{5}, we present an SSW in a diffraction grating with
$Q\sim1/\delta^8$.
It is obtained using the direct computation method developed in the same
section.

For an ASW, $f_j$ is even in $y$ for odd $j$ and odd in $y$ for even $j$.
Hence, $u_1$ is even in $y$ and generally radiates to infinity, giving
$d_1\neq0$ and $Q\sim1/\delta^2$.
If $d_1=0$, the ASW becomes a super-ASW.
In this case, the radiation coefficients vanish at all even orders, and
the leading radiation order satisfies
$\mu=3,5,7,\ldots$.
The case $\mu=3$ has been studied in many works
~\cite{EBulgakov2017PRA,LYuan2020PRAPerturbation,EBulgakov2023PRB,NZhang2024OL,LLiu2024OE}.
Super-ASWs with $\mu=5$ and $7$ were first reported in
Refs.~\cite{HLuo2023PRA,LLiu2024OE}, where they were obtained by tuning two or
three structural parameters to merge BICs on a
single dispersion curve, respectively.
In Sec.~\Rmnum{5}, we present a numerical example of a super-ASW with
$\mu=5$ and locate it using the direct computation method.

For an off-$\Gamma$ BIC, the field has no definite parity in $y$, 
and hence neither does $f_1$.
The first field correction $u_1$ therefore generally radiates to infinity, 
and an off-$\Gamma$ BIC is a super-BIC only if $d_1=0$.
More generally, no parity constraint restricts $f_j$ to be even or odd, 
so an off-$\Gamma$ super-BIC can in principle have any $\mu\ge 2$.
Examples with $\mu=2$ and $\mu=3$ can be found in Refs.~\cite{EBulgakov2017PRA,LLiu2024OE}.

Finally, we consider degenerate BICs.
For simplicity, we focus on doubly degenerate at-$\Gamma$ BICs with two
orthonormal states $\phi_1$ and $\phi_2$.
If both $\phi_1$ and $\phi_2$ are SSWs, Eq.~\eqref{matrixA} gives
${\bm A}=0$ and hence $\omega_1=0$.
Since $u_*$ is a linear combination of $\phi_1$ and $\phi_2$, it is also
even in $y$, and the first-order source $f_1$ is odd.
The radiating part of $u_1$ is therefore odd and has vanishing radiation.
Thus, $d_1=0$, and the degenerate BIC is a super-BIC.

If both $\phi_1$ and $\phi_2$ are ASWs, we again have
${\bm A}=0$ and $\omega_1=0$.
In this case, $u_*$ is odd in $y$, and hence $f_1$ is even.
The radiating part of $u_1$ is therefore even and generally has a
nonzero radiation coefficient.
Thus, $d_1$ is generally nonzero, and the degenerate BIC is generally
not a super-BIC.

If $\phi_1$ and $\phi_2$ have opposite parities in $y$ but the same
parity in $z$, the diagonal elements of ${\bm A}$ vanish, while the
off-diagonal elements are generally nonzero.
Equation~\eqref{degenerateomega1} then gives two opposite nonzero values
of $\omega_1$, corresponding to two resonant bands with linear
dispersion near $\Gamma$.
The degenerate BIC therefore forms a Dirac point.
Each corresponding $u_*$ is a linear combination of an SSW and an ASW
and thus has no definite parity in $y$.
Consequently, the first-order radiation is generally nonzero, and the
Dirac-point BIC is generally not a super-BIC.

\section{Unfolding of super-BICs under structural perturbations}

In Secs.~\Rmnum{2} and \Rmnum{3}, we analyze the radiation
behavior of resonant states near BICs and classify
super-BICs by $\mu$ and symmetry.
In this section, we investigate the unfolding of super-BICs
under structural perturbations~\cite{MGolubitsky1985Book,YKuznetsov2004Book}.
Since super-BICs are generally non-generic,
structural variations may turn them into generic BICs~\cite{NZhang2024OE,NZhang2024OL}.
We first develop a general bifurcation theory
to characterize the behavior of nearby generic BICs.
We then establish a theory of parametric dependence
to determine the minimum number $n$ of additional
tunable structural parameters required to recover
a nearby super-BIC of the same class.
This number also gives the minimum number of
tunable structural parameters required to realize that class.
Building on previous studies of the parametric dependence
of generic BICs and the bifurcations of certain super-BIC
classes~\cite{LYuan2020PRAParametric,NZhang2024OE,NZhang2024OL},
we develop a unified theory for nondegenerate super-BICs.
The framework can also be extended to degenerate cases.

\subsection{Bifurcation theory}

Let $(\beta_*,\omega_*,u_*)$ be a nondegenerate super-BIC
with $\mu\geq2$. We consider the structural variation
\begin{equation}
	\label{Case1perdie}
	\varepsilon({\bm r};\eta)
	=\varepsilon_*({\bm r})+\eta F({\bm r}),
\end{equation}
where $\eta$ is a small real parameter and $F({\bm r})$
is real, periodic, and invariant under reflections
in $y$ and $z$.
We first examine the resonant band containing the
original super-BIC and its dependence on $\eta$.
Let $\delta=(\beta-\beta_*)L$.
In a neighborhood of $(\delta,\eta)=(0,0)$,
the outgoing radiation coefficient of nearby
resonant states can be expanded as
\begin{equation}
	\label{eq:sec4_dexp}
	d(\delta,\eta)
	=\sum_{j=0}^\infty d_j(\eta)\delta^j,\quad
	d_j(\eta)
	=\sum_{k=0}^\infty d_{jk}\eta^k,
\end{equation}
where $d_j(0)=0$ for $j<\mu$ and
$d_\mu(0)\ne0$~\cite{BZhen2014PRL}.
We investigate the existence of nearby generic BICs
$(\beta,\omega,u)$ satisfying $d=0$ in the perturbed
structure and their dependence on $\eta$.

We first consider SSWs and off-$\Gamma$ super-BICs.
In the unperturbed structure, a wavenumber
detuning $\delta$ produces a leading radiation
amplitude $d_\mu\delta^\mu$.
Under a structural variation, the radiation coefficient
at $\delta=0$ generally becomes nonzero, with
$d_0(\eta)\sim d_{01}\eta$,
where $d_{01}\ne0$ for a generic structural variation.
For a nearby BIC, these two contributions cancel
to leading order, giving
\begin{equation}
	\label{eq:sec4_leading_generic}
	\delta^\mu-C\eta=0,\quad
	C:=-\frac{d_{01}}{d_\mu},
\end{equation}
which suggests the scaling
$\eta={\cal O}(\delta^\mu)$.

For a super-ASW, reflection symmetry preserves the
symmetry-protected BIC at $\delta=0$ under structural
variations, so that $d_0(\eta)\equiv0$.
Under a generic structural variation, the coefficient
$d_1(\eta)$ satisfies $d_1(\eta)\sim d_{11}\eta$,
where $d_{11}\ne0$.
For nearby BICs with $\delta\ne0$, balancing
the leading contributions yields
\begin{equation}
	\label{eq:sec4_leading_asw}
	\delta^{\mu-1}-C\eta=0,\quad
	C:=-\frac{d_{11}}{d_\mu},
\end{equation}
which suggests the scaling
$\eta={\cal O}(\delta^{\mu-1})$.

The above leading-order conditions indicate the possible
locations of nearby BICs, satisfying
$\delta^p\sim C\eta$, where $p=\mu$ for SSWs
and off-$\Gamma$ super-BICs, and $p=\mu-1$
for super-ASWs with $\delta\ne0$.
These relations motivate a Puiseux-series ansatz
for nearby BICs.
We develop below a bifurcation theory to formally
construct these BIC branches and establish their
bifurcation behaviors.
In particular, we show that $C$ is real and that,
under generic structural perturbations with $C\ne0$,
SSWs and off-$\Gamma$ super-BICs with even $\mu$
exhibit saddle-node-type bifurcations, with two
nearby BIC branches for $C\eta>0$ and none
for $C\eta<0$.
For super-ASWs with odd $\mu$, the symmetry-protected
BIC persists, while two additional off-$\Gamma$
BIC branches emerge for $C\eta>0$, producing
a pitchfork-type bifurcation.
For off-$\Gamma$ super-BICs with odd $\mu$,
a single BIC branch continues through the
super-BIC for either sign of $\eta$,
without bifurcation.
These bifurcation behaviors are numerically
verified in Sec.~\Rmnum{6}.

We seek nearby BICs admitting the following
Puiseux expansions for their Bloch wavenumbers
and frequencies:
\begin{equation}
	\label{eq:sec4_betaw_exp}
	\begin{bmatrix}
		\beta\\
		\omega
	\end{bmatrix}
	=
	\begin{bmatrix}
		\beta_*\\
		\omega_*
	\end{bmatrix}
	+\sum_{j=1}^\infty\eta^{j/p}
	\begin{bmatrix}
		\beta_j\\
		\omega_j
	\end{bmatrix},
\end{equation}
with the corresponding field expanded as
\begin{equation}
	\label{eq:sec4_u_exp}
	u({\bm r};\eta)
	=e^{i[\beta(\eta)-\beta_*]y}
	\sum_{j=0}^\infty\eta^{j/p}u_j({\bm r}),
	\quad u_0=u_*.
\end{equation}
The correction fields have the same quasi-periodicity
as $u_*$.
Substitution into the Helmholtz equation gives
\begin{equation}
	\label{eq:sec4_field_order}
	{\cal L}_*u_j=g_j,\quad j\geq1.
\end{equation}
The right-hand side $g_j$ is obtained from the wavevector
and frequency expansions, together with the
structural perturbation $\eta F({\bm r})$.
For $j<p$, $g_j$ can be expressed
as a linear combination of $f_1,\ldots,f_j$
in Eq.~\eqref{EqLj}.
For $j\ge p$, additional contributions arise
from the structural perturbation.
If there exist real expansion coefficients in
Eq.~\eqref{eq:sec4_betaw_exp} such that
Eq.~\eqref{eq:sec4_field_order} admits nonradiating
solutions at every order, we obtain a formal
construction of nearby BIC branches.

The solvability conditions for a nonradiating solution
$u_j$ are
\begin{equation}
	\label{eq:sec4_conditions}
	\langle u_*|g_j\rangle=0,\quad
	\langle v_s|g_j\rangle=0,
\end{equation}
where $v_s$ is a scattering state of the same up-down
parity as the BIC, satisfying the gauge condition
$\langle v_s|\varepsilon_*|u_*\rangle=0$.
The first condition ensures Fredholm solvability
as in Sec.~\Rmnum{2}, while the second eliminates
outgoing radiation, as follows from Green's identity.
We introduce the operator ${\cal C}$, defined by
\begin{equation}
	\label{eq:sec4_C}
	({\cal C}w)(y,z)=\overline{w}(-y,z).
\end{equation}
The phases of $u_*$ and $v_s$ can be chosen such that
${\cal C}u_*=u_*$ and ${\cal C}v_s=v_s$.
Suppose that the lower-order expansion coefficients
are real and the corresponding fields are nonradiating
and ${\cal C}$-invariant.
For real $\beta_j$ and $\omega_j$, we then have
${\cal C}g_j=g_j$.
Consequently, both conditions in
Eq.~\eqref{eq:sec4_conditions} are real.
If these conditions admit real solutions for the
expansion coefficients, the corresponding nonradiating
field $u_j$ can also be chosen ${\cal C}$-invariant.
This property allows the construction to proceed
inductively, provided the real solvability conditions
can be satisfied at every order.

For the nearby off-$\Gamma$ BIC branches arising
from the super-BICs considered here, solving
Eq.~\eqref{eq:sec4_field_order} order by order
shows that the second condition in
Eq.~\eqref{eq:sec4_conditions} is automatically
satisfied for $j<\mu$.
At $j=\mu$, this condition yields the real equation
\begin{equation}
	\label{eq:sec4_beta1}
	\beta_1^p=-\frac{B_F}{A_*L^p},
\end{equation}
where $A_*\ne0$ is proportional to the leading
radiation coefficient $d_\mu$ of the original
super-BIC and is independent of $F({\bm r})$,
while $B_F$ depends linearly on $F({\bm r})$.	
It follows that the coefficient $C$ defined in Eqs.~\eqref{eq:sec4_leading_generic}--\eqref{eq:sec4_leading_asw}
satisfies $C=-B_F/A_*$.
Generic structural perturbations
correspond to $B_F\ne0$.
Once a real root of Eq.~\eqref{eq:sec4_beta1} is selected,
the solvability conditions can be satisfied recursively
at higher orders, yielding a formal construction of
nearby BIC branches.

\subsection{Parametric dependence theory}

The preceding analysis shows that a generic structural
perturbation generally destroys the original super-BIC,
although nearby generic BICs may remain.
We investigate whether the super-BIC can be recovered
by tuning additional structural parameters.
We consider
\begin{equation}
	\label{newperdie}
	\varepsilon({\bm r};\eta,{\bm\xi})
	=\varepsilon_*({\bm r})+\eta F({\bm r})
	+\sum_{\ell=1}^n\xi_\ell G_\ell({\bm r}),
\end{equation}
where ${\bm\xi}=[\xi_1,\ldots,\xi_n]^{\sf T}$
contains the tunable structural parameters.
The profiles $G_\ell({\bm r})$ are real, periodic,
and reflection symmetric, as is $F({\bm r})$.

We establish two results.
First, a nearby super-BIC with $\mu\ge2$ can
generically be recovered using one additional
tuning parameter, although its value of $\mu$
may change.
Second, if the original $\mu$ must be preserved,
the minimum number $n$ of tuning parameters is
$n=\mu/2$ for SSWs, $n=(\mu-1)/2$ for super-ASWs,
and $n=\mu-1$ for off-$\Gamma$ super-BICs.
In both cases, the tuning parameters can be chosen
such that ${\bm\xi}(\eta)={\cal O}(\eta)$.
The results for prescribed $\mu$ are summarized
in Table~\Rmnum{1}, where the entries in red
are verified numerically in Sec.~\Rmnum{5}.
\begin{table}[h]
	\centering
	\caption{
		Minimum number $n$ of tunable structural
		parameters required to preserve a super-BIC
		with prescribed $\mu$.
		Red entries are verified numerically.
	}
	\label{Table1}
	\begin{tabular}{|c|c|c|c|}
		\hline
		$\mu$ & SSW & super-ASW & off-$\Gamma$ super-BIC\\
		\hline
		$2$ & \textcolor{red}{\textbf{1}}
		& -- & \textcolor{red}{\textbf{1}} \\
		\hline
		$3$ & --
		& \textcolor{red}{\textbf{1}}
		& \textcolor{red}{\textbf{2}} \\
		\hline
		$4$ & \textcolor{red}{\textbf{2}}
		& -- & $3$ \\
		\hline
		$5$ & --
		& \textcolor{red}{\textbf{2}}
		& $4$ \\
		\hline
		$6$ & $3$ & -- & $5$ \\
		\hline
		$7$ & -- & $3$ & $6$ \\
		\hline
		$\vdots$ & $\vdots$ & $\vdots$ & $\vdots$ \\
		\hline
	\end{tabular}
\end{table}

We first consider the recovery of a super-BIC with
$\mu\ge2$, without specifying its exact value.
At the original super-BIC, the field $u_*$
and its first wavevector correction $u_1$
satisfy
\begin{equation}
	\label{paradepMain}
	{\cal L}_*{\bm u}_*
	={\bm f}_*,\quad
	{\bm u}_*=
	\begin{bmatrix}u_*\\u_1\end{bmatrix},\quad
	{\bm f}_*=
	\begin{bmatrix}0\\f_1\end{bmatrix},
\end{equation}
where ${\bm u}_*\to0$ as $z\to\pm\infty$.
We seek a nearby super-BIC
$(\tilde\beta_*,\tilde\omega_*,\tilde u_*)$
whose corresponding field vector
$\tilde{\bm u}_*$ satisfies
\begin{equation}
	\label{PerMainEq}
	\tilde{\cal L}_*
	\tilde{\bm u}_*
	=\tilde{\bm f}_*.
\end{equation}
We introduce the expansions
\begin{equation}
	\label{SuperBICAsb1}
	\begin{bmatrix}
		\tilde\beta_*\\
		\tilde\omega_*\\
		\tilde\omega_1
	\end{bmatrix}
	=
	\sum_{j=0}^\infty\eta^j
	\begin{bmatrix}
		\beta_{*,j}\\
		\omega_{*,j}\\
		\omega_{1,j}
	\end{bmatrix},
\end{equation}
and
\begin{equation}
	\label{eq:sec4_param_field}
	{\bm\xi}=\sum_{j=1}^\infty\eta^j{\bm\xi}_j,
	\quad
	\tilde{\bm u}_*
	=e^{i(\tilde\beta_*-\beta_*)y}
	\sum_{j=0}^\infty\eta^j{\bm u}_j,
\end{equation}
where ${\bm u}_0={\bm u}_*$.

Substituting these expansions into
Eq.~\eqref{PerMainEq} gives, at order $\eta^j$,
\begin{equation}
	\label{PerMainEqj}
	{\cal L}_*{\bm u}_j
	={\bm f}_j({\bm r};\beta_{*,j},
	\omega_{*,j},\omega_{1,j},{\bm\xi}_j).
\end{equation}
As in the preceding subsection, we use the
${\cal C}$ symmetry to obtain real solvability
conditions and construct nonradiating
field corrections.
Suppose that the lower-order expansion coefficients
are real and the corresponding fields are
nonradiating and ${\cal C}$-invariant.
For real $j$-th order coefficients, the source
${\bm f}_j$ is also ${\cal C}$-invariant.
The Fredholm and nonradiation conditions
for the two components therefore give four
real equations, linear in the $j$-th order
coefficients:
\begin{equation}
	\label{LinearEquationParaDep}
	{\bm M}
	\begin{bmatrix}
		\beta_{*,j}\\
		\omega_{*,j}\\
		\omega_{1,j}\\
		{\bm\xi}_j
	\end{bmatrix}
	={\bm b}_j,
\end{equation}
where ${\bm M}$ is a $4\times(3+n)$ real matrix,
and ${\bm b}_j$ depends on $F({\bm r})$ and the
lower-order solutions.
If these equations admit real solutions,
the corresponding nonradiating field corrections
can also be chosen ${\cal C}$-invariant.
Thus, the construction can proceed recursively,
provided the real solvability conditions
can be satisfied at every order.

We write the coefficient matrix as
\begin{equation}
	\label{eq:sec4_M}
	{\bm M}
	=
	\left[
	{\bm M}_*\;
	{\bm m}(G_1)\;
	\cdots\;
	{\bm m}(G_n)
	\right],
\end{equation}
where ${\bm M}_*$ is the $4\times3$ matrix
associated with $\beta_{*,j}$, $\omega_{*,j}$,
and $\omega_{1,j}$, while each
${\bm m}(G_\ell)$ is the column associated
with the tuning parameter $\xi_\ell$.
Generically, $\operatorname{rank}({\bm M}_*)=3$.
Without additional tuning parameters,
solvability requires
${\bm q}^{\sf T}{\bm b}_j=0$,
where ${\bm q}$ is a nonzero left null vector
of ${\bm M}_*$.
This condition is not generically satisfied
for a prescribed perturbation $F({\bm r})$.
Introducing one tuning parameter $\xi_1$
with a generic profile $G_1({\bm r})$
supplies a fourth independent column,
making ${\bm M}$ nonsingular.
The real solvability conditions can then
be satisfied recursively at every order,
yielding a formal construction of a nearby
super-BIC with $\mu\ge2$, although its
original value of $\mu$ may change.
This establishes our first result.

We next require the original value of $\mu$
to remain unchanged.
In this case, all field corrections
$u_1,\ldots,u_{\mu-1}$ must remain nonradiating.
We therefore extend the field vector to
$[u_*,u_1,\ldots,u_{\mu-1}]^{\sf T}$
and include the frequency coefficients
$\omega_2,\ldots,\omega_{\mu-1}$
in Eq.~\eqref{SuperBICAsb1}.
The same analysis gives $2\mu$ real Fredholm
and nonradiation conditions at each order.

For an off-$\Gamma$ super-BIC, the unknowns
at each order are $\beta_{*,j}$,
$\omega_{*,j},\omega_{1,j},\ldots,
\omega_{\mu-1,j}$, and ${\bm\xi}_j$.
The coefficient matrix ${\bm M}$ has size
$2\mu\times(\mu+1+n)$.
The submatrix ${\bm M}_*$ formed by the first
$\mu+1$ columns generically has full column rank.
Each independent tuning profile $G_\ell({\bm r})$
increases the rank by one until full row rank
is reached.
Therefore, the minimum number of tuning
parameters is $n=\mu-1$.
For at-$\Gamma$ super-BICs, reflection symmetry
in $y$ eliminates the wavevector corrections
and odd-indexed frequency coefficients,
leaving $\mu$ real equations.
For an SSW with even $\mu$, the coefficients
$\omega_{*,j},\omega_{2,j},\ldots,
\omega_{\mu-2,j}$ provide $\mu/2$
independent columns, giving $n=\mu/2$.
For a super-ASW with odd $\mu$, the coefficients
$\omega_{*,j},\omega_{2,j},\ldots,
\omega_{\mu-1,j}$ provide $(\mu+1)/2$
independent columns, giving $n=(\mu-1)/2$.
With these numbers of independent tuning parameters,
the solvability conditions can be satisfied
recursively at higher orders, yielding a formal
construction of nearby super-BICs with
the original value of $\mu$.

Let ${\cal P}$ be an $m$-dimensional structural
parameter space.
The minimum number $n$ is the codimension
of a given super-BIC class in ${\cal P}$.
Consequently, super-BICs of this class
generically form an $(m-n)$-dimensional set
in ${\cal P}$.
For example, super-ASWs with $\mu=3$ form curves
in a 2D parameter space, whereas those with
$\mu=5$ occur at isolated points.
These predictions facilitate the direct computation
of super-BICs in Sec.~\Rmnum{5}.

\section{Direct Computation of Super-BICs and Numerical Examples}
\begin{figure*}[htbp]
	\centering
	\includegraphics[scale=0.33]{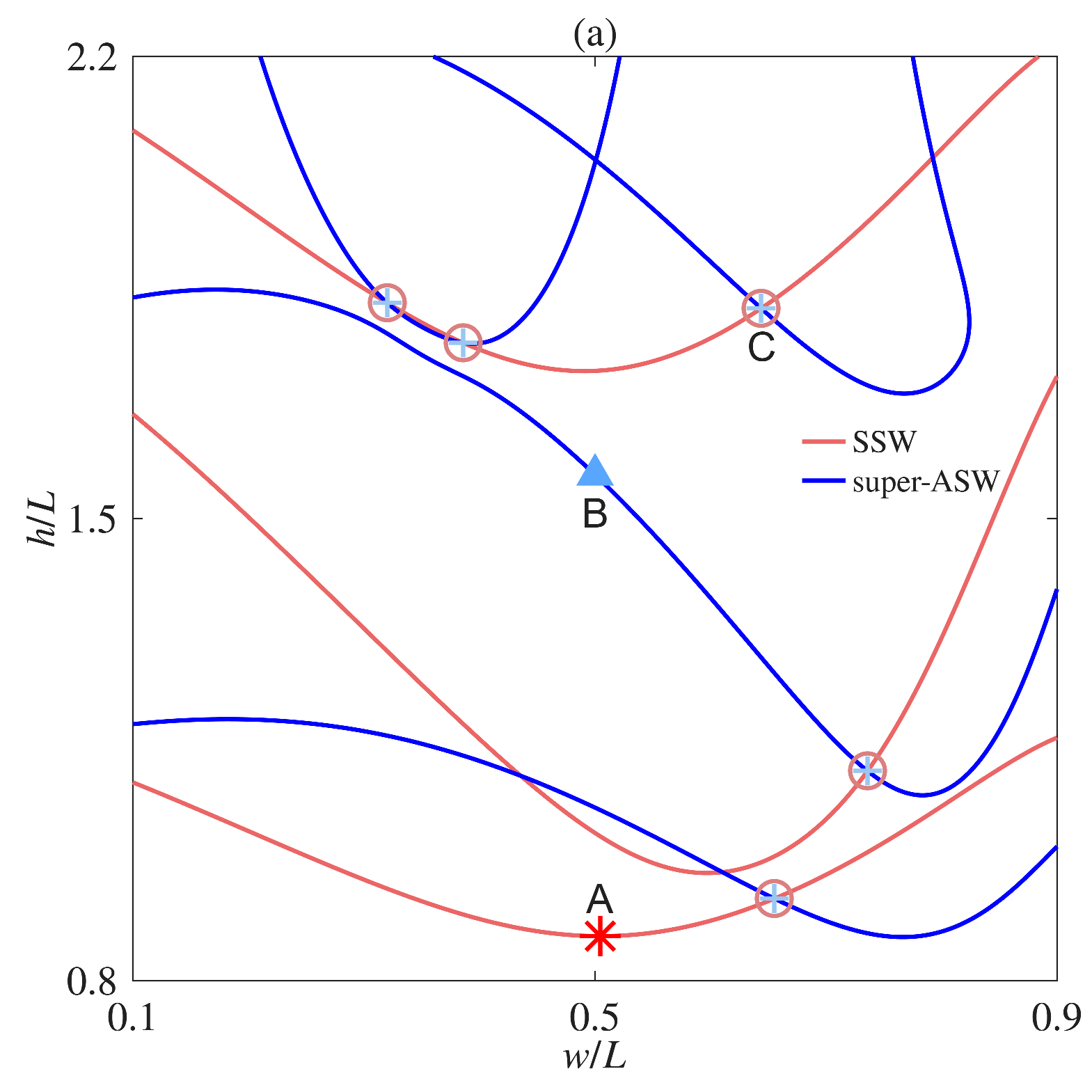}
	\hspace{0.2cm}
	\includegraphics[scale=0.33]{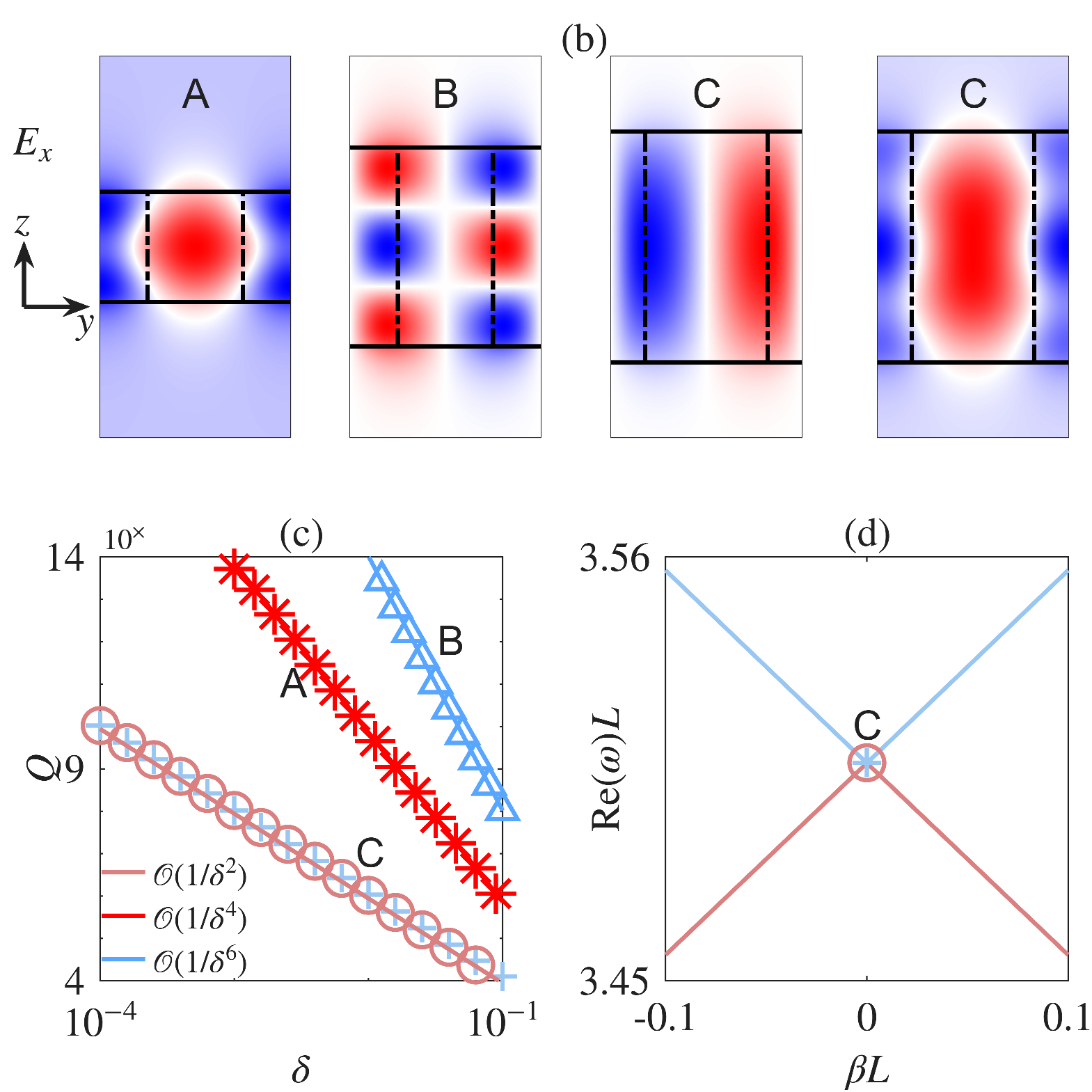}
	\caption{
		At-$\Gamma$ super-BICs, i.e., SSWs and super-ASWs,
		in a diffraction grating shown in
		Fig.~\ref{structure}(a) with $n_1=1.45$ and $n_2=2.55$.
		(a) SSWs (red lines) and super-ASWs (blue lines)
		in the parameter space $(w,h)$.
		(b) The electric field component $E_x$ for
		$\sf A$ (SSW), $\sf B$ (super-ASW), and
		$\sf C$ (degenerate BIC).
		(c) $Q$ factors of resonant states near these
		three BICs.
		The scalings $Q\sim1/\delta^2$,
		$Q\sim1/\delta^4$, and $Q\sim1/\delta^6$
		hold for $\sf C$, $\sf A$, and $\sf B$,
		respectively.
		(d) $\operatorname{Re}(\omega)$ of resonant states
		near the degenerate BIC $\sf C$ (a Dirac point).
	}
	\label{superswwh}
\end{figure*}
\begin{table*}[htbp]
	\centering
	\caption{
		Parameters of the labeled BICs used in the numerical examples.
		The type specifies the BIC class, while $\mu$ denotes
		the leading radiation order of nearby resonances.
	}
	\label{tab:bic_parameters}
	{\setlength{\tabcolsep}{8pt}
		\renewcommand{\arraystretch}{1.08}
		\begin{tabular}{c c c c c c c c c}
			\hline\hline
			Label & Type & $\mu$ & $n_1$ & $n_2$ & $w/L$ & $h/L$
			& $\omega L$ & $\beta L$ \\
			\hline
			\BICasterisk\,{\sf A}
			& SSW & 2 & 1.45 & 2.55
			& 0.5060 & 0.8676 & 3.685 & 0
			\\
			\BICpoly{3}{markerB}{0}\,{\sf B}
			& super-ASW & 3 & 1.45 & 2.55
			& 0.5000 & 1.567 & 3.713 & 0
			\\
			\BICdouble\,{\sf C}
			& degenerate BIC & 1 & 1.45 & 2.55
			& 0.6439 & 1.819 & 3.507 & 0
			\\
			\BICpoly{4}{markerD}{45}\,{\sf D}
			& super-ASW & 5 & 1.45 & 2.55
			& 0.4057 & 1.697 & 3.390 & 0
			\\
			\BICstarPoly{5}{markerE}\,{\sf E}
			& off-$\Gamma$ super-BIC & 2 & 1.45 & 2.55
			& 0.5502 & 1.545 & 3.793 & 0.7340
			\\
			\BICstarPoly{6}{markerE}\,{\sf F}
			& off-$\Gamma$ super-BIC & 3 & 2.55 & 3.5
			& 0.3279 & 1.860 & 4.085 & 0.3184
			\\
			\BICcross{markerG}\,{\sf G}
			& SSW & 4 & 2.55 & 3.5
			& 0.3393 & 1.838 & 4.134 & 0
			\\
			\BICpoly{4}{markerH}{0}\,{\sf H}
			& super-ASW & 3 & 1.45 & 2.55
			& 0.8236 & 1.798 & 4.265 & 0
			\\
			\hline\hline
	\end{tabular}}
\end{table*}

In the preceding sections, we showed that 
a super-BIC with $Q\sim Q_{{2{\mu}}}/\delta^{{2{\mu}}}$, ${\mu}\geq 2$, 
can be identified by the conditions $d_j=0$ for $j<\mu$ and ${d}_{\mu}\neq 0$.
We also determined the minimum number $n$ of tunable structural
parameters required to find a super-BIC. 
Using these results, in this section,
we propose a computation procedure to directly identify super-BICs
together with the corresponding structural parameters. 
Numerical examples are presented to illustrate our method. 

\subsection{Direct computation}

Let ${\bm p}=(p_1,\ldots,p_m)\in{\cal P}$ denote the structural
parameters. For trial real values of $\omega$, $\beta$, and ${\bm p}$,
we consider the auxiliary eigenvalue problem
\begin{align}
	\label{eq:aux_eig}
	&{\cal L}(\omega,{\bm p})u_0 = \lambda_0 u_0,\\
	\label{eq:aux_bloch}
	&u_0(y+L,z) = e^{i\beta L}u_0(y,z),
\end{align}
with the outgoing radiation condition.
Here ${\cal L}$ is the Helmholtz operator in
Eq.~\eqref{mainEq}, with dielectric function
$\varepsilon({\bm r};{\bm p})$.
We use the perturbation equations of Sec.~\Rmnum{2},
replacing $u_*$ with $u_0$:
\begin{equation}
	\label{eq:aux_corr}
	({\cal L}_0-\lambda_0)u_j
	=f_j({\bm r};\omega_j), \quad j\geq 1,
\end{equation}
where $\omega_j$ is an undetermined coefficient and $u_j$
is an unknown function satisfying the same quasi-periodicity
as $u_0$. When $\lambda_0=0$, the field $u_0$ is a BIC,
and $\omega_j$ and $u_j$ become the corresponding
frequency and field corrections.

Numerically, the auxiliary eigenvalue problem and the perturbation
equations are discretized as
\begin{align}
	\label{eq:aux_discrete}
	&{\bm  L}(\lambda_0;\omega,\beta,{\bm p}){\bm  u}_0
	={\bm  0},\\
	\label{eq:aux_discrete_corr}
	&{\bm  L}(\lambda_0;\omega,\beta,{\bm p}){\bm  u}_j
	={\bm  f}_j,\quad j\geq 1.
\end{align}
The matrix ${\bm  L}$ can depend linearly on $\lambda_0$,
for example in finite-element or finite-difference formulations
with perfectly matched layers, or nonlinearly, as in methods
based on transverse impedance operators~\cite{NZhang2026JCP}
or the Fourier modal method~\cite{NZhang2026arXiv}.
At each order, $\omega_j$ is determined by the solvability
condition of the discretized perturbation equation,
and $u_j$ is then obtained with an appropriate gauge condition.

The outgoing radiation coefficients $d_j$ are extracted
from the field corrections $u_j$.
A super-BIC with prescribed $\mu$ can therefore be located by solving
\begin{equation}
	\label{eq:direct_aux_residual}
	{\bm G}(\omega,\beta,{\bm p})=
	\begin{bmatrix}
		\lambda_0\\
		d_1\\
		\vdots\\
		d_{\mu-1}
	\end{bmatrix}
	={\bm 0},
\end{equation}
and verifying that $d_\mu\neq0$.
The nonlinear system can be solved using Newton's method,
and super-BIC families can be traced by continuation.

Alternatively, a nested iterative scheme can be used.
For each trial ${\bm p}$, the inner iteration locates a BIC
by solving $\lambda_0(\omega,\beta;{\bm p})=0$
for $\omega({\bm p})$ and $\beta({\bm p})$.
The radiation coefficients $d_j({\bm p})$ are then evaluated
at this BIC. The outer iteration adjusts ${\bm p}$ until
\begin{equation}
	\label{eq:nestedmethod}
	{\bm G}({\bm p})=
	\begin{bmatrix}
		d_1\\
		\vdots\\
		d_{\mu-1}
	\end{bmatrix}
	={\bm 0}.
\end{equation}

Since all structures considered in this work are layered
in $z$, we evaluate ${\bm G}$
using a method based on the Fourier modal method
~\cite{NZhang2026arXiv}.
In the following, we calculate super-BICs in the
diffraction gratings depicted in Fig.~\ref{structure}(a).

\subsection{Numerical Examples: At-$\Gamma$ super-BICs}

We consider a diffraction grating embedded in air. 
The grating consists of two materials with refractive indices $n_1$ and $n_2$,
where $n_1=1.45$ for silicon dioxide (${\rm SiO}_2$) and $n_2=2.55$ for titanium dioxide (${\rm TiO}_2$) in the optical spectrum.
Using the method outlined above, 
we first compute at-$\Gamma$ super-BICs, including SSWs and super-ASWs, in the parameter space ${\cal P}=(w,h)$.
In Fig.~\ref{superswwh}(a), we mark several families of super-ASWs and
SSWs that are even in $z$ by blue and red lines, respectively. 
To verify the results, for super-BICs {\sf A} and {\sf B} indicated in Fig.~\ref{superswwh}(a), 
we plot their electric fields and show the $Q$ factors of nearby resonant states in Fig.~\ref{superswwh}(b) and (c), respectively.
The numerical parameters of {\sf A}, {\sf B}, and all other labeled super-BICs
are listed in Table~\ref{tab:bic_parameters}.

From Fig.~\ref{superswwh}(a), it is clear that the curves of
super-ASWs and SSWs intersect at several points.  
It turns out that the points marked by red circles and blue plus signs
correspond to doubly degenerate BICs, each comprising an SSW and an ASW.
Other intersection points correspond to two BICs with different
frequencies and are thus non-degenerate. 
For the degenerate BIC labeled $\sf C$, 
we plot its electric field components $E_x$ in Fig.~\ref{superswwh}(b), 
plot the $Q$ factor and  $\mbox{Re}(\omega)$ of nearby resonant states in Fig.~\ref{superswwh}(c) and (d),
respectively.
As shown in Fig.~\ref{superswwh}(d), the degenerate BIC can be regarded as a Dirac point,
characterized by a locally linear variation in $\omega$ of nearby resonant states: $\omega\approx \omega_*\pm\delta\omega_1$. 
However, Fig.~\ref{superswwh}(c) shows that the
$Q$ factor in both the upper and lower bands follows $Q\sim
1/\delta^2$. Therefore, this degenerate BIC is in fact not a super-BIC.
This finding is notably counterintuitive. The degenerate BIC lies at
the intersection of two super-BIC parameter curves, but does not yield
ultrahigh-$Q$ resonances for nearby Bloch wavenumber. 

\subsection{Numerical Examples: Off-$\Gamma$ super-BICs}

\begin{figure}[htbp]
	\centering 
	\includegraphics[scale=0.33]{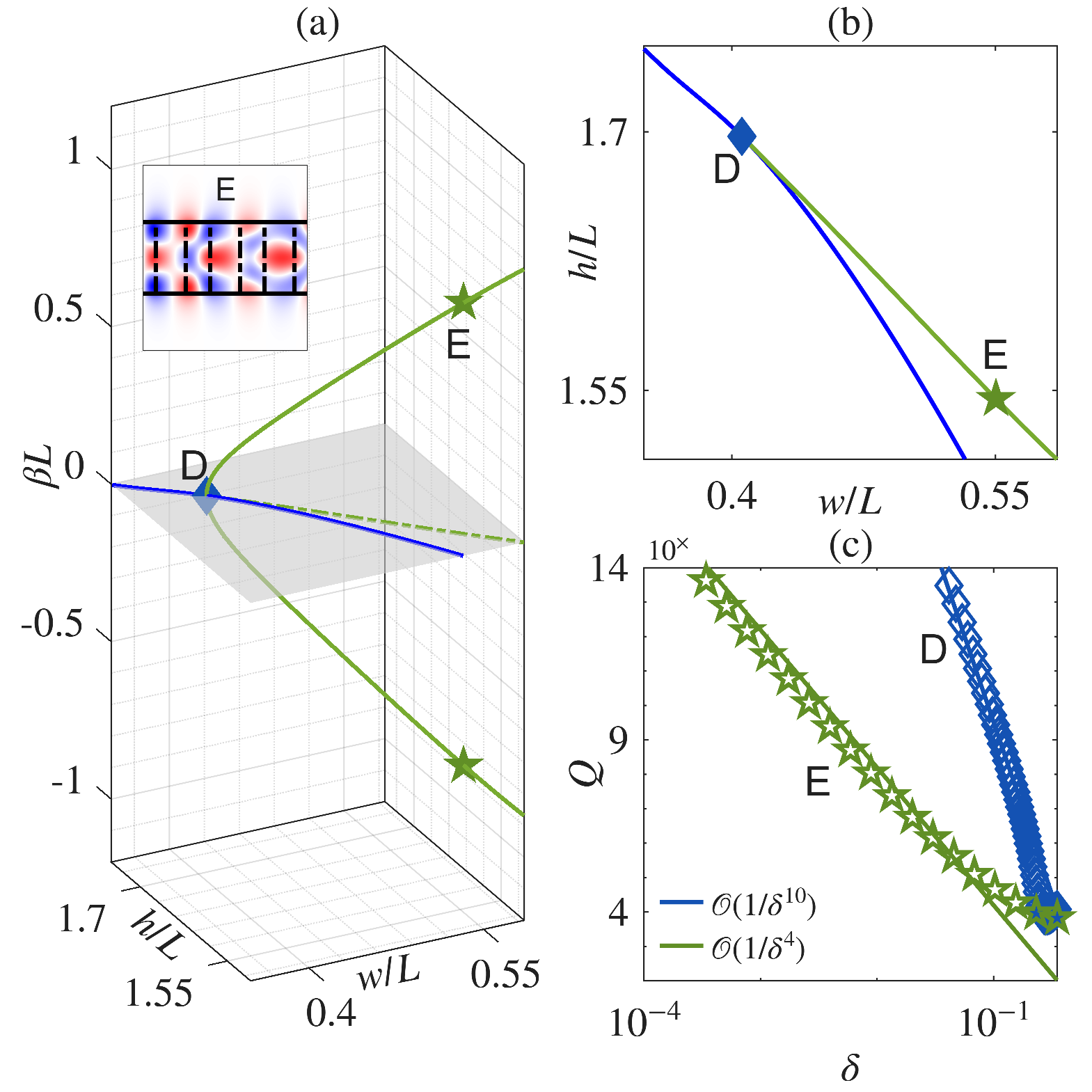}
	\caption{(a). The values of $(w,h,\beta)$ of super-BICs.
		The inset shows the field component $\mbox{Re}(E_x)$ of super-BIC {\sf E}.
		(b). The parameter values supporting super-BICs form curves in $(w,h)$ space.
		(c). $Q$ factors of resonant states near super-BICs {\sf D} and {\sf E}. 
		It is clear that $Q\sim 1/\delta^{10}$ holds for super-BIC $\sf D$.}\label{3Doffgammasuperbic}
\end{figure}

For the same grating structure, we compute off-$\Gamma$ super-BICs in $(w,h)$ space and show the corresponding values of $(w,h,\beta)$ in Fig.~\ref{3Doffgammasuperbic}(a).
The green solid curve represents off-$\Gamma$ super-BICs, 
while its projection onto the plane $\beta=0$ (the gray plane) is indicated by the green dashed line. 
As a verification, 
for an off-$\Gamma$ super-BIC labeled as {\sf E},
we plot the real part of its field component $E_x$ in the inset of Fig.~\ref{3Doffgammasuperbic}(a),
and plot the $Q$ factor of nearby resonant states in Fig.~\ref{3Doffgammasuperbic}(c).

\begin{figure}[htbp]
	\centering 
	\includegraphics[scale=0.33]{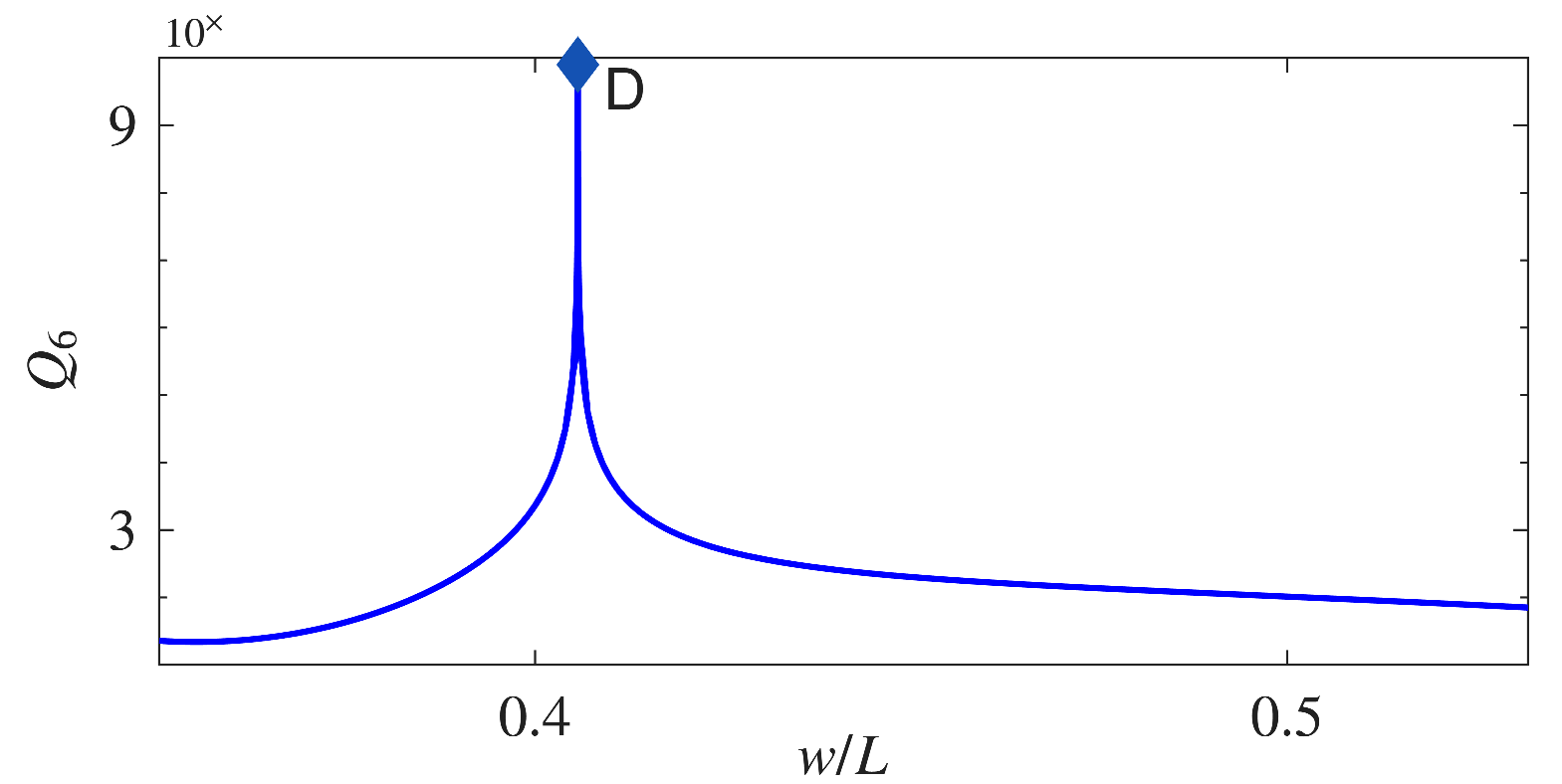}
	\caption{The asymptotic coefficients $Q_6$ for super-ASWs illustrated in Fig.~\ref{3Doffgammasuperbic}. 
		The blue diamond represents the super-BIC {\sf D}.}\label{Q6nearsupersuper}
\end{figure}

Notably, as shown in Fig.~\ref{3Doffgammasuperbic}(a), 
an intersection between the super-ASWs (the blue line) and the off-$\Gamma$ super-BICs curves occurs at point {\sf D}.
For clarity, we show parameter values supporting these super-BICs in Fig.~\ref{3Doffgammasuperbic}(b). 
The curves for the super-ASWs and the
off-$\Gamma$ super-BICs form a cusp at {\sf D}.
The super-ASWs on the blue curve in Fig.~\ref{3Doffgammasuperbic} typically
exhibit $Q\sim Q_6/\delta^6$, but {\sf D} is a special super-ASW with ${\mu}=5$.
As shown in Fig.~\ref{3Doffgammasuperbic}(c), the $Q$ factor of resonant
states near {\sf D} follows $Q\sim 1/\delta^{10}$. 
We also show the
coefficient $Q_6$ in Fig.~\ref{Q6nearsupersuper}. 
It is clear that $Q_6$ tends to infinity at {\sf D}, 
leading to a jump in the asymptotic order from 6 to 10. 

If the usual merging process is used to find such a super-ASW with $\mu=5$,
one first identifies an ASW and four off-$\Gamma$ BICs lying on the
same dispersion curve.
As will be elaborated in Sec.~\Rmnum{6},
this configuration occurs only when $(w,h)$ lies within the narrow region
between the blue and green curves in Fig.~\ref{3Doffgammasuperbic}(b).
The structural parameters $(w,h)$ must then be tuned continuously toward
${\sf D}$, where all five BICs coalesce into a single BIC.
In contrast, our direct method is far more efficient: we simply solve the
nonlinear system ${\bm G}(\omega,\beta,{\bm p})=0$, where
${\bm p}=(w,h)\in{\cal P}$.
During the iterations, ${\bm p}$ need not remain within the narrow region,
nor is it necessary to track the individual BICs. 

\begin{figure}[htbp]
	\centering 
	\includegraphics[scale=0.33]{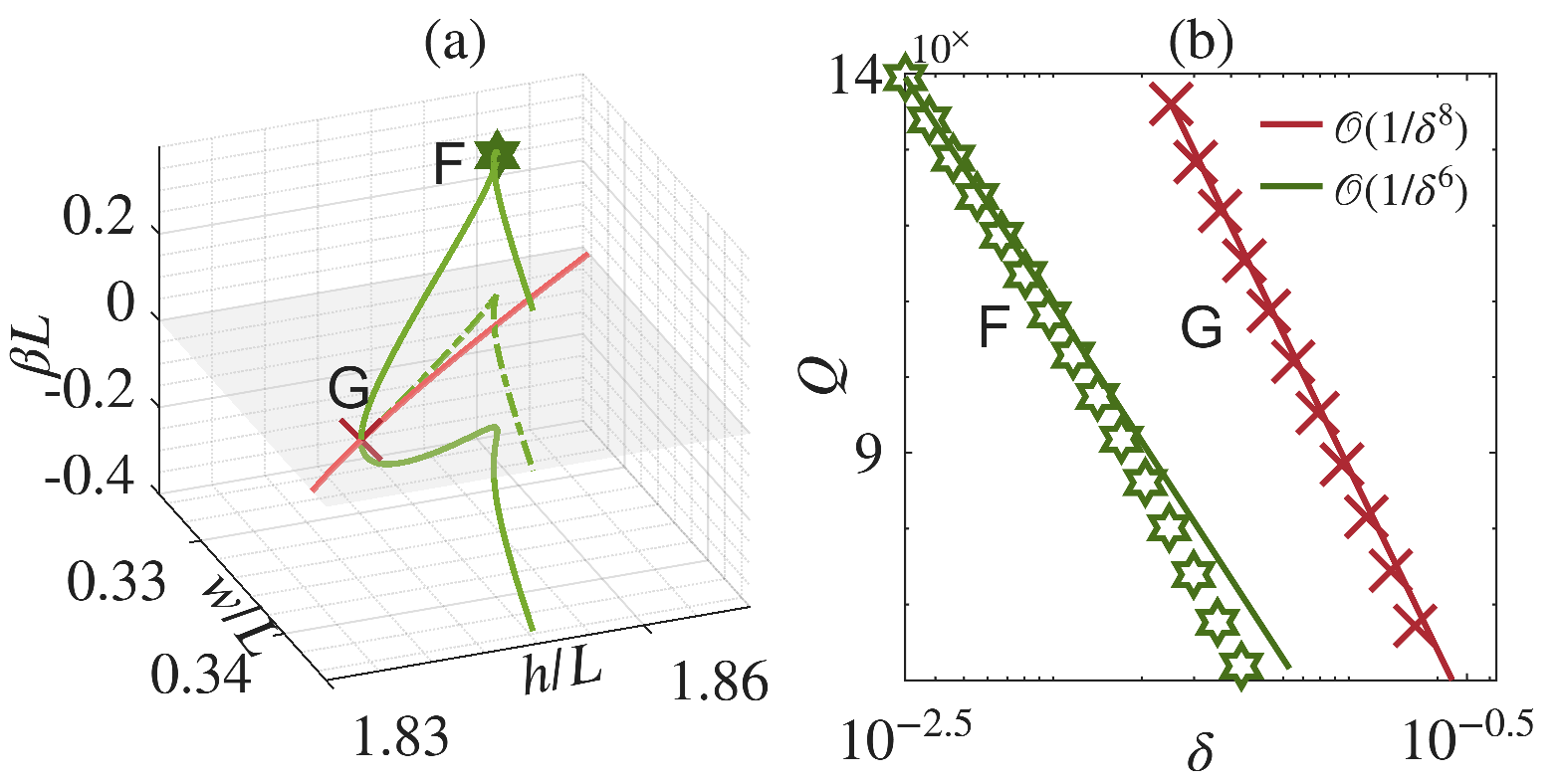}
	\caption{(a). The values of $(w,h,\beta)$ of super-BICs.
		(b). $Q$ factor of resonant states near super-BICs {\sf F} and {\sf G}.}\label{NX3line}
\end{figure}
To see some different super-BICs, we consider another grating structure with
refractive indices $n_1=2.55$ and $n_2=3.5$ (silicon in the optical spectrum). 
We calculate some super-BICs and show their values of
$(w,h,\beta)$ in Fig.~\ref{NX3line}(a). 
The red curve represents a family of SSWs. It can be observed that the
two parameter curves (for 
the off-$\Gamma$ super-BICs and the SSWs, respectively) intersect at point
{\sf G} marked by a red cross.  
From Fig.~\ref{NX3line}(b), it is clear that 
the $Q$ factor of resonant states near SSW ${\sf G}$ follows $Q\sim
1/\delta^8$. This implies that ${\sf G}$ is an at-$\Gamma$ super-BIC with ${\mu}=4$. 
In addition, on the curve for off-$\Gamma$ super-BICs, there is a 
special one with ${\mu}=3$ and it is labeled as {\sf F} and marked by a green pentagon. 
As shown in Fig.~\ref{NX3line}(b), the $Q$ factor for resonant states
near ${\sf F}$ exhibits the asymptotic relation $Q\sim 1/\delta^6$.

\subsection{Numerical Examples: Super-BICs with higher values of ${\mu}$}
\begin{figure}[htbp]
	\centering 
	\includegraphics[scale=0.33]{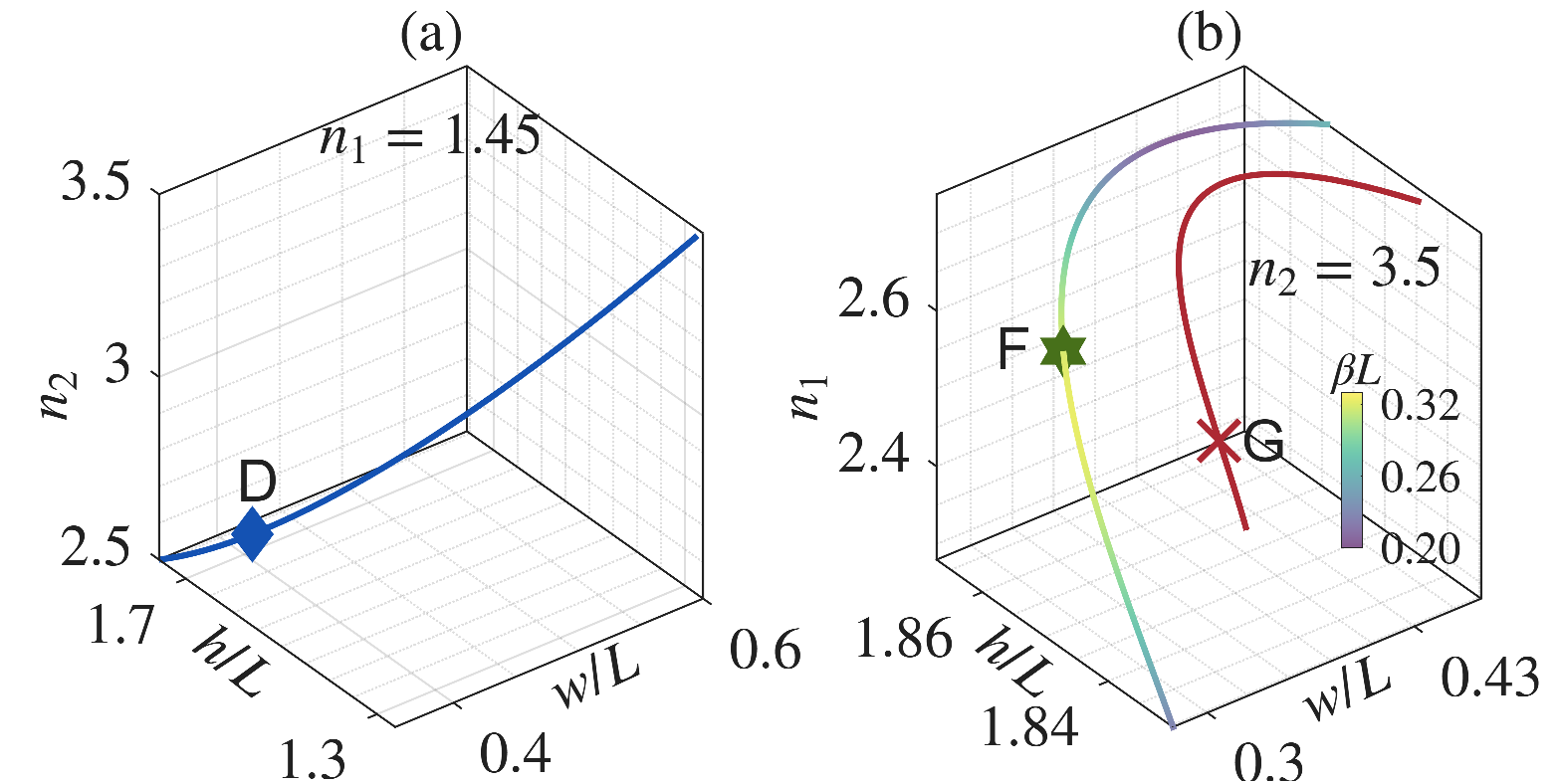}
	\caption{(a). Parameter values of $(w,h,n_2)$ with a fixed $n_1=1.45$ that support super-ASWs with ${\mu}=5$.
		(b). Parameter values of $(w,h,n_1)$ with a fixed $n_2=3.5$ that
		support SSWs with ${\mu}=4$ (blue) and off-$\Gamma$ super-BICs with
		${\mu}=3$ (color change curve).}
	\label{ContinuousTwoParameters}
\end{figure}
So far, we have found that SSWs with $\mu=2$,
super-ASWs with $\mu=3$, and off-$\Gamma$
super-BICs with $\mu=2$ form curves in the structural parameter space $(w,h)$,
whereas the super-ASW ${\sf D}$ with $\mu=5$,
SSW ${\sf G}$ with $\mu=4$, and off-$\Gamma$
super-BIC ${\sf F}$ with $\mu=3$ occur at isolated points.
The three super-BICs with larger values of $\mu$
all have $n=2$, consistent with our theory of
parametric dependence summarized in Table~\Rmnum{1}.
We next show that each of them forms a continuous
curve in a three-dimensional structural parameter space.
For simplicity, we use the refractive index $n_1$ or $n_2$
as the third parameter.
For the super-ASW ${\sf D}$ with $\mu=5$,
we fix $n_1=1.45$ and consider the parameter space
$(w,h,n_2)$.
For super-BICs ${\sf F}$ and ${\sf G}$,
we fix $n_2=3.5$ and consider the parameter space
$(w,h,n_1)$.
The numerical results are shown in
Fig.~\ref{ContinuousTwoParameters},
where ${\sf D}$, ${\sf F}$, and ${\sf G}$ are highlighted.
The off-$\Gamma$ super-BIC curve in
Fig.~\ref{ContinuousTwoParameters}(b) is color-coded
by the Bloch wavenumber $\beta$.

\section{Geometry of super-BICs in parameter space}
\label{sec:bifurcation}

In Sec.~\Rmnum{4}, we studied the unfolding of super-BICs
under structural perturbations, including the bifurcations
of nearby generic BICs and the recovery of super-BICs
by tuning additional structural parameters.
These results admit a unified geometric interpretation
in the product space ${\cal P}\times\mathrm{BZ}$,
where ${\cal P}$ denotes the structural parameter space
and $\mathrm{BZ}$ the first Brillouin zone.
We denote by ${\cal B}$ the collection of BICs in this
product space, characterized by the nonradiation
condition $d=0$.
The number $n$ is the codimension of the corresponding
super-BIC submanifold within the BIC manifold.
A structural path
${\bm p}(\eta)\in{\cal P}$ selects a section of
${\cal B}$ that describes the bifurcations of nearby
generic BICs.

\begin{figure}[htbp]
	\centering
	\includegraphics[scale=0.33]{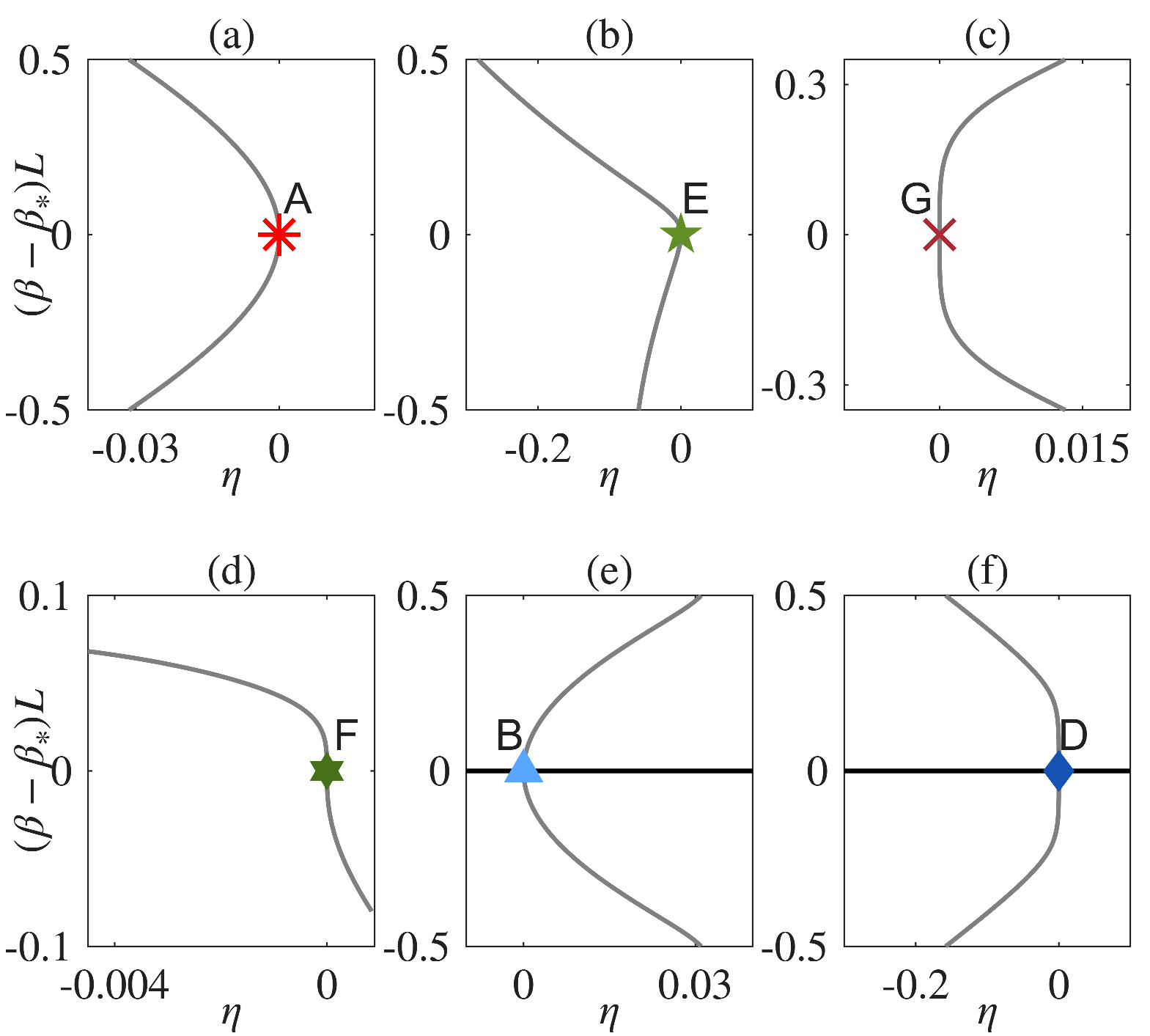}
	\caption{
		BIC branches along generic structural paths
		through the super-BICs obtained in
		Sec.~\Rmnum{5}.
		The structural parameters of the super-BICs
		are denoted by $(w_l,h_l)$, with $w=w_l$
		fixed and $\eta=(h-h_l)/L$.
		(a),(b) Saddle-node bifurcations associated
		with the $\mu=2$ SSW {\sf A} and
		off-$\Gamma$ super-BIC {\sf E}, respectively.
		(c) Higher-order saddle-node-type bifurcation
		at the $\mu=4$ SSW {\sf G}.
		(d) Cubic-root continuation through the
		$\mu=3$ off-$\Gamma$ super-BIC {\sf F},
		without bifurcation.
		(e) Pitchfork bifurcation at the
		$\mu=3$ super-ASW {\sf B}.
		(f) Higher-order pitchfork bifurcation
		at the $\mu=5$ super-ASW {\sf D}.
	}
	\label{bifurcation}
\end{figure}

We first numerically verify the bifurcation behaviors
predicted in Sec.~\Rmnum{4} along generic structural paths.
Figures~\ref{bifurcation}(a)--(c) show
saddle-node-type bifurcations at
{\sf A}, {\sf E}, and {\sf G}, including
the higher-order scaling
$\eta\sim \delta^4$ at {\sf G},
where $\delta=(\beta-\beta_*)L$.
Figure~\ref{bifurcation}(d) shows the
continuation through {\sf F} without bifurcation.
Figures~\ref{bifurcation}(e) and (f)
show pitchfork-type bifurcations at
{\sf B} and {\sf D}, respectively,
with $\eta\sim \delta^4$
at {\sf D}.

The commonly studied BIC-merging process can be
viewed as the reverse of certain bifurcations
shown in Fig.~\ref{bifurcation}, such as those
at {\sf A}, {\sf B}, and {\sf E}~\cite{NZhang2024OL}.
However, $\mu$ alone does not determine
the number of merging BIC branches.
For example, only two off-$\Gamma$ BIC branches
merge with the $\mu=4$ SSW {\sf G}, while the
$\mu=3$ off-$\Gamma$ super-BIC {\sf F}
does not involve any BIC merging.

The preceding examples describe bifurcations along
particular structural paths, but do not reveal
the full organization of nearby BICs in
${\cal P}\times\mathrm{BZ}$.
To understand how the bifurcation behavior depends
on the choice of structural path, we examine
the local geometry of ${\cal B}$ near super-BICs.
For concreteness, we consider a 2D structural
parameter space and a 1D Brillouin zone,
focusing on quadratic and cusp geometries.

\subsection{Quadratic geometries}

Consider a super-BIC at
$(\delta,\eta_1,\eta_2)=(0,0,0)$, where $\eta_1$ and $\eta_2$
are local structural coordinates. Suppose that a local BIC surface
can be written as $\eta_2=g(\eta_1,\delta)$. We focus on a critical
point of this surface, for which
$\partial_{\eta_1}g=\partial_\delta g=0$. The leading local geometry
is then determined by the quadratic expansion
\begin{equation}
	\eta_2
	\sim
	\begin{bmatrix}
		\eta_1 & \delta
	\end{bmatrix}
	{\bm K}
	\begin{bmatrix}
		\eta_1\\
		\delta
	\end{bmatrix}, \quad
	{\bm K}
	=
	\begin{bmatrix}
		a & c\\
		c & b
	\end{bmatrix},
	\label{eq:quadratic_geometry}
\end{equation}
where
\begin{equation}
	a=\frac{1}{2}\partial_{\eta_1}^2g,
	\quad
	c=\frac{1}{2}\partial_{\eta_1\delta}g,
	\quad
	b=\frac{1}{2}\partial_\delta^2g.
	\label{eq:quadratic_matrix}
\end{equation}
The matrix ${\bm K}$ is real symmetric, and the sign of
$\det{\bm K}$ distinguishes two local geometries.
For $\det{\bm K}<0$, the BIC surface has a saddle point,
while for $\det{\bm K}>0$, it has an elliptic point.
The singular case $\det{\bm K}=0$ is not considered here
but can be analyzed similarly.

\begin{figure}[htbp]
	\centering
	\includegraphics[scale=0.33]{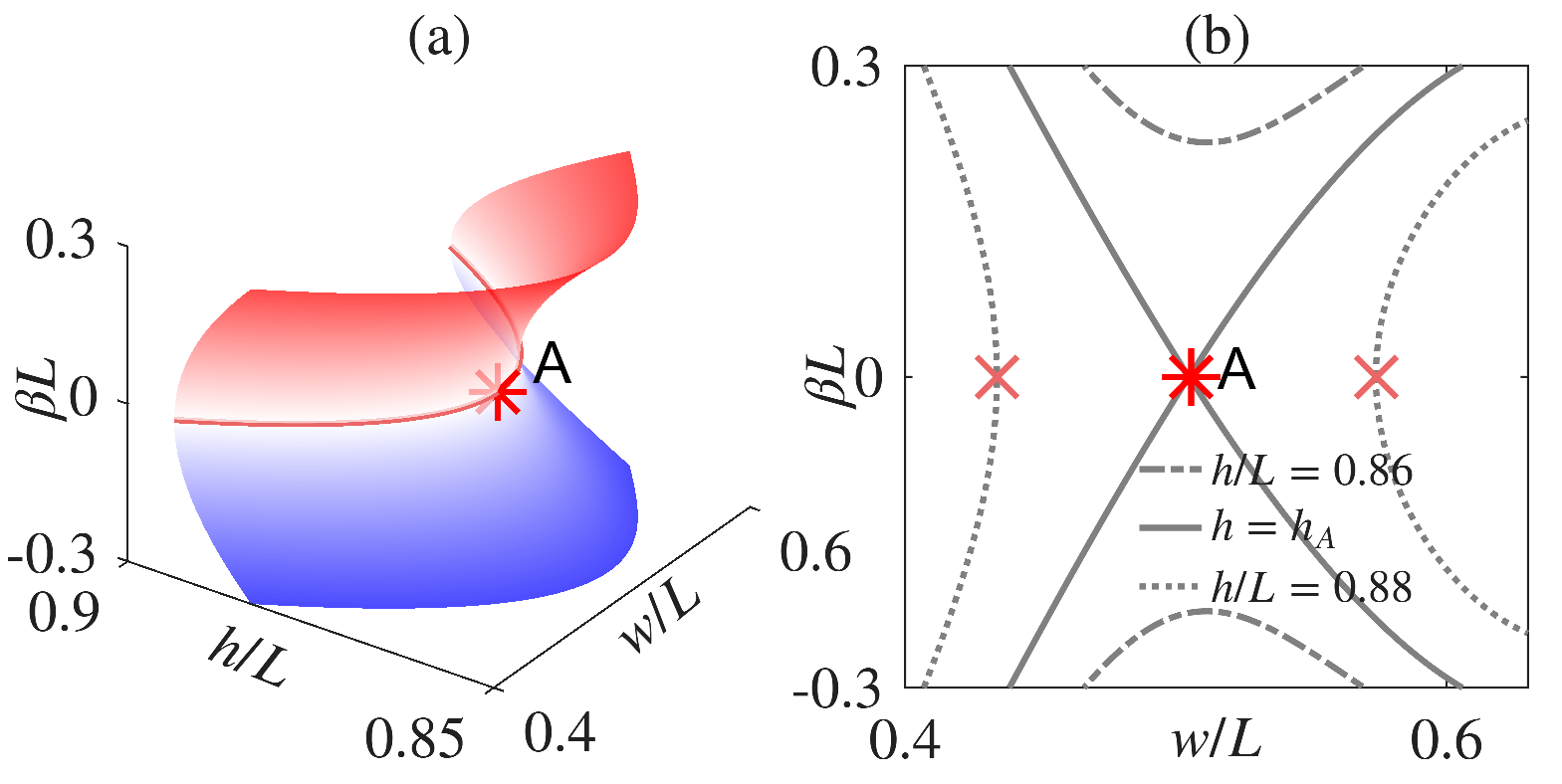}
	\caption{
		Local BIC geometry near SSW {\sf A}.
		(a) BICs in $(w,h,\beta)$ space, with {\sf A} at the
		saddle point.
		(b) Sections at fixed values of $h$. At $h=h_A$, the
		section consists locally of two linear branches.
	}
	\label{sswA_bifurcation}
\end{figure}

The BIC set near SSW {\sf A} provides an example of
the saddle geometry, as shown in
Fig.~\ref{sswA_bifurcation}(a).
In this BIC manifold, the red curve is the branch
with $\delta=0$, corresponding to a family of SSWs.
Here we take
$\eta_1=(w-w_A)/L$ and $\eta_2=(h-h_A)/L$.
At {\sf A}, $\partial_\delta d=0$, while numerical calculations yield
$\partial_{\eta_1}d=0$ and $\partial_{\eta_2}d\neq0$.
The leading expansion of 
$d(\delta,\eta_1,\eta_2)$ near {\sf A} is therefore
\begin{equation}
	d
	\sim
	(\partial_{\eta_2}d)\big|_{\sf A}\eta_2
	+
	\frac{1}{2}(\partial_{\eta_1}^2d)\big|_{\sf A}\eta_1^2
	+
	\frac{1}{2}(\partial_\delta^2d)\big|_{\sf A}\delta^2.
	\label{eq:A_taylor}
\end{equation}
The implicit-function theorem allows the BIC surface $d=0$
to be written locally as
$\eta_2=g(\eta_1,\delta)$.
Differentiating this condition and evaluating at
${\sf A}$, we obtain
\begin{equation}
	\left.\partial_{\eta_1}g\right|_{\sf A}
	=
	-\left.
	\frac{\partial_{\eta_1}d}{\partial_{\eta_2}d}
	\right|_{\sf A}
	=0,
	\quad
	\left.\partial_\delta g\right|_{\sf A}
	=
	-\left.
	\frac{\partial_\delta d}{\partial_{\eta_2}d}
	\right|_{\sf A}
	=0.
\end{equation}
Thus, ${\sf A}$ is a critical point of the BIC surface.
Solving Eq.~\eqref{eq:A_taylor} for $\eta_2$ gives
Eq.~\eqref{eq:quadratic_geometry}, with
\begin{equation}
	a=-\frac{\partial_{\eta_1}^2d}{2\partial_{\eta_2}d},
	\quad
	c=0,
	\quad
	b=-\frac{\partial_\delta^2d}{2\partial_{\eta_2}d}.
	\label{eq:A_quadratic_coefficients}
\end{equation}
At {\sf A}, numerical evaluation gives $a>0$ and $b<0$,
so $\det{\bm K}=ab<0$.
Therefore, {\sf A} is a saddle point, with the
local BIC surface given by
$\eta_2\sim a\eta_1^2+b\delta^2$.

For a fixed nonzero $\eta_2$, the BIC section is locally a
hyperbola with two branches. As $\eta_2$ passes through zero, the
orientation of the hyperbola changes. At the critical section
$\eta_2=0$, $a\eta_1^2+b\delta^2=0$, and the two branches satisfy
\begin{equation}
	\delta
	\sim
	\pm\sqrt{-\frac{a}{b}}\,\eta_1,
	\label{eq:saddle_linear}
\end{equation}
so that they intersect linearly at {\sf A}. This behavior is shown by
the fixed-$h$ sections in Fig.~\ref{sswA_bifurcation}(b).

Different structural paths through {\sf A} exhibit different
bifurcation behaviors. Along $w=w_A$, $\eta_1=0$, and
$\partial_{\eta_2}d\neq0$, so this is a generic structural path.
The local relation $\eta_2\sim b\delta^2$ produces the saddle-node
bifurcation shown in Fig.~\ref{bifurcation}(a).
Along $h=h_A$, $\eta_2=0$. Since $\partial_{\eta_1}d=0$ at {\sf A},
this path is non-generic, and the BIC branches instead undergo
the linear bifurcation described by Eq.~\eqref{eq:saddle_linear}.
Finally, the section $\delta=0$ gives the SSW family,
locally described by $\eta_2\sim a\eta_1^2$.
Along a structural path following this family, the BIC remains
at $\delta=0$, and no bifurcation occurs locally.

\begin{figure}[htbp]
	\centering
	\includegraphics[scale=0.33]{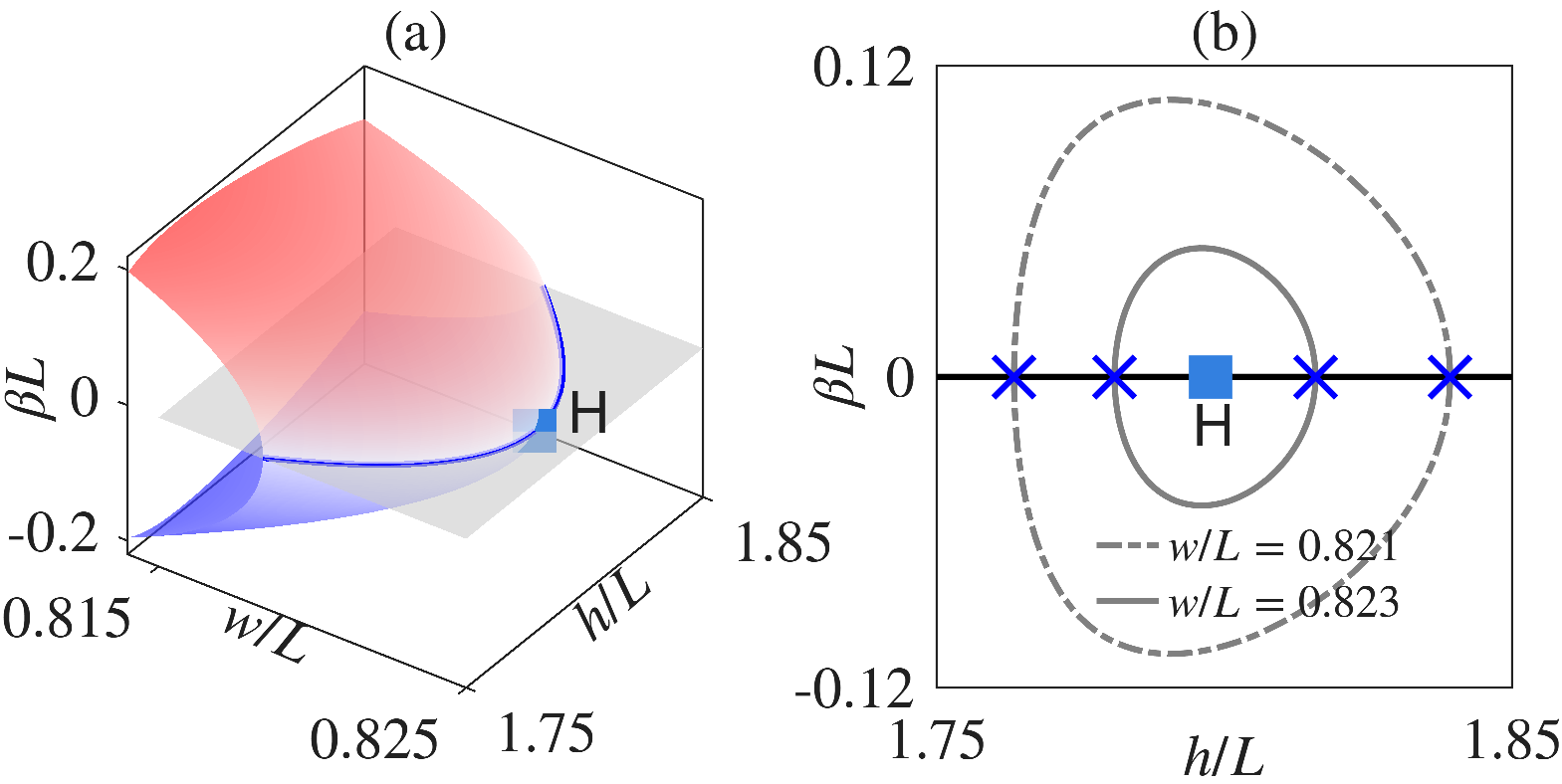}
	\caption{
		Local BIC geometry near the super-ASW {\sf H}.
		(a) Off-$\Gamma$ BICs in $(w,h,\beta)$ space near {\sf H}.
		The gray plane represents the symmetry-protected ASWs.
		(b) Sections at fixed values of $w$. The closed BIC loops
		contract toward {\sf H} as $w$ approaches $w_H$.
	}
	\label{superaswH_geometry}
\end{figure}

We next consider the BIC set near the super-ASW
{\sf H}, shown in Fig.~\ref{superaswH_geometry}(a).
The symmetry-protected ASWs form a BIC surface
at $\delta=0$.
Another surface consists of nearby off-$\Gamma$
BICs and a family of super-ASWs, represented by
the blue curve in the figure.
The two surfaces intersect along this super-ASW family.
We focus on the second surface, which has a local
elliptic geometry.

Here we take
$\eta_1=(h-h_H)/L$ and $\eta_2=(w-w_H)/L$.
Numerical calculations give
$\partial_{\eta_1}d_1=0$ and
$\partial_{\eta_2}d_1\neq0$ at {\sf H}.
Thus, the implicit-function theorem allows the
second BIC surface to be written locally as
$\eta_2=g(\eta_1,\delta)$, with {\sf H} as a
critical point.
Numerical evaluation gives $a<0$ and $b<0$,
while reflection symmetry implies $c=0$.
Therefore, $\det{\bm K}=ab>0$,
confirming that {\sf H} is an elliptic point.

The local BIC surface is given by
$\eta_2\sim a\eta_1^2+b\delta^2$.
For fixed $\eta_2<0$, corresponding to $w<w_H$,
the BIC section is locally a closed ellipse,
while no nearby off-$\Gamma$ BIC exists
for $\eta_2>0$.
As $\eta_2$ approaches zero, the ellipse
contracts to {\sf H}, as shown in
Fig.~\ref{superaswH_geometry}(b).

Different structural paths through {\sf H}
produce different bifurcation behaviors.
Along $\eta_1=0$, $\eta_2$ varies, and
$\partial_{\eta_2}d_1\neq0$, so this is a
generic structural path.
The local relation $\eta_2\sim b\delta^2$,
together with the symmetry-protected BIC
branch $\delta=0$, produces a pitchfork bifurcation.
Along $\eta_2=0$, $\eta_1$ varies.
Since $\partial_{\eta_1}d_1=0$ at {\sf H},
this path is non-generic.
In this case, $a\eta_1^2+b\delta^2=0$
has no nonzero real solutions because $ab>0$.
Thus, only the symmetry-protected BIC branch remains.
This behavior has been reported previously~\cite{NZhang2024OL},
but its interpretation in terms of the local BIC geometry
was not discussed.
Finally, the super-ASW family is locally described
by $\eta_2\sim a\eta_1^2$ at $\delta=0$.
Along a structural path following this family,
only the super-ASWs branch remains,
and no bifurcation occurs locally.

\subsection{Cusp geometry}

\begin{figure}[htbp]
	\centering
	\includegraphics[scale=0.33]{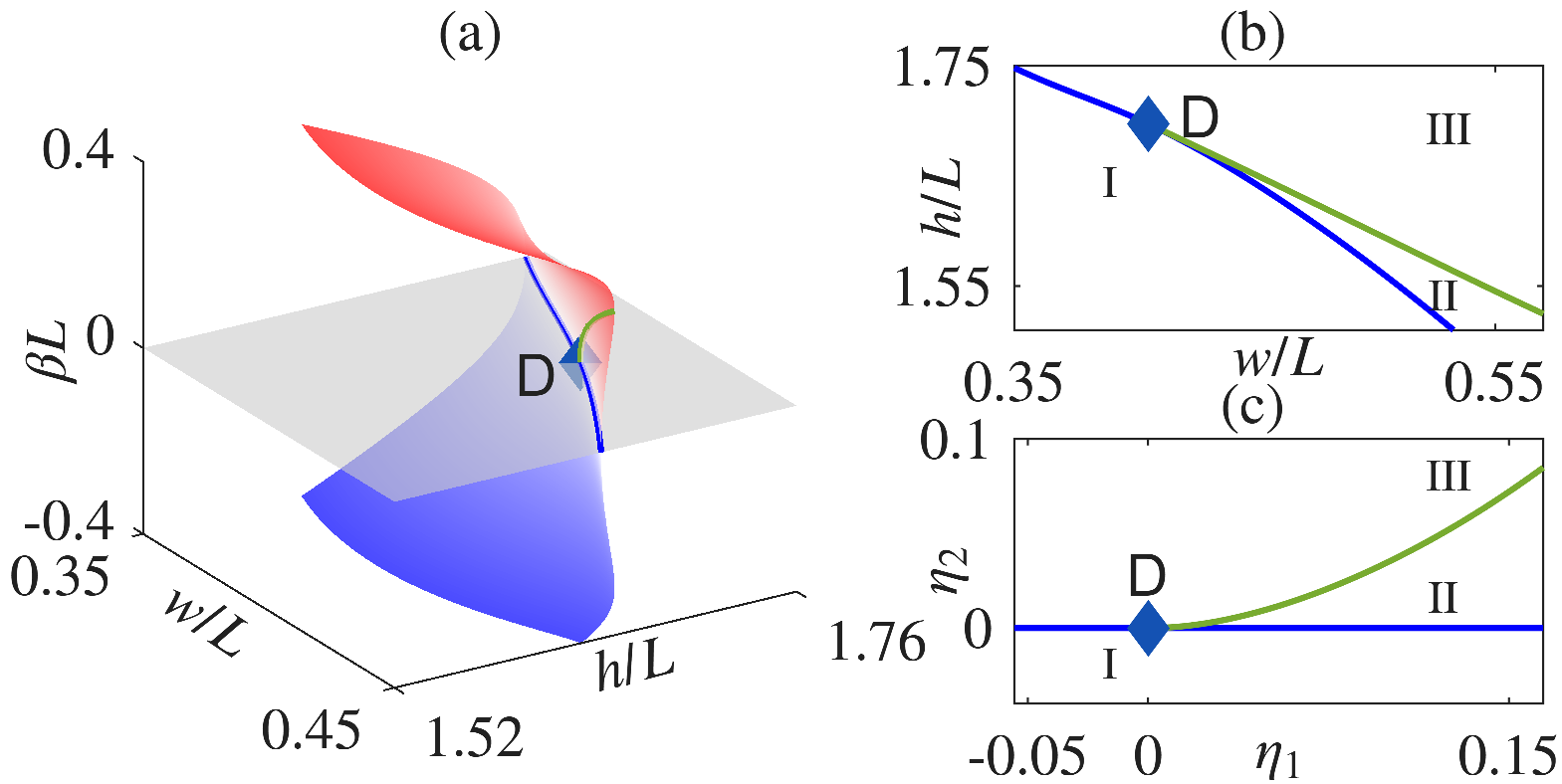}
	\includegraphics[scale=0.33]{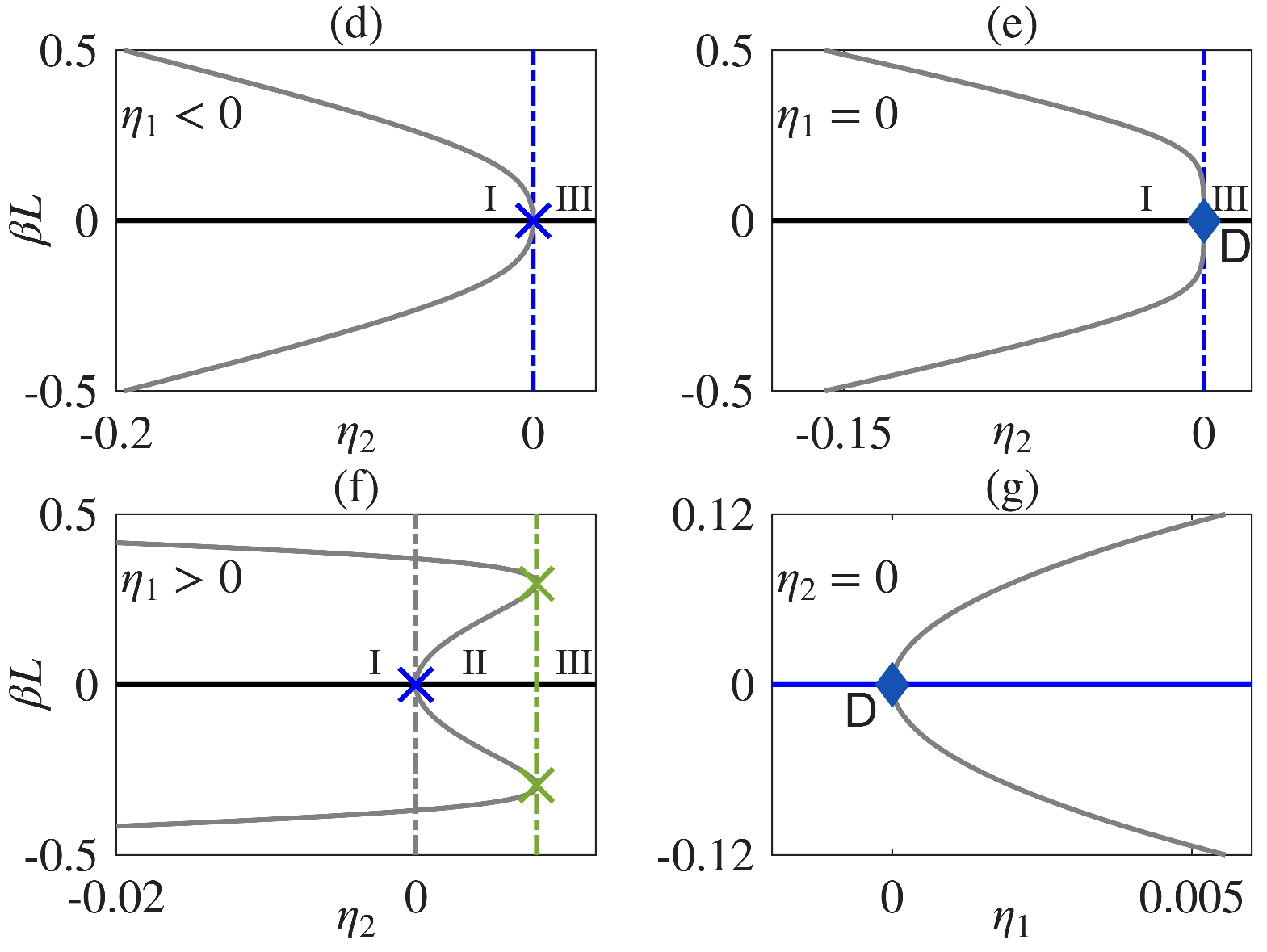}
	\caption{
		Local BIC geometry near the $\mu=5$ super-ASW {\sf D}.
		(a) BICs in $(w,h,\beta)$ space near {\sf D}.
		(b) The super-ASW and off-$\Gamma$ super-BIC families in
		the $(w,h)$ plane.
		(c) The same families in $(\eta_1,\eta_2)$ coordinates,
		dividing the local parameter space into regions I--III.
		(d)--(f) BIC branches as functions of $\eta_2$ for fixed
		$\eta_1<0$, $\eta_1=0$, and $\eta_1>0$, respectively.
		(g) BIC branches along the super-ASW family $\eta_2=0$.
	}
	\label{superaswDgeometry}
\end{figure}

We next examine a cusp-like geometry near the
$\mu=5$ super-ASW {\sf D}.
As shown in Figs.~\ref{3Doffgammasuperbic}
and \ref{superaswDgeometry}(a), the super-ASW
family and the off-$\Gamma$ super-BIC branches
meet at {\sf D}, forming a cusp-like structure
in the local BIC set.
Related singular geometries also arise in
catastrophe theory~\cite{VArnold1986Book}.

As shown in Fig.~\ref{superaswDgeometry}(b), the super-ASW family
locally forms a smooth curve. We write it as $h=h_*(w)$, with
$h_*(w_D)=h_D$, and introduce the local coordinates
\begin{equation}
	\eta_1=\frac{w-w_D}{L},
	\quad
	\eta_2=\frac{h-h_*(w)}{L}.
	\label{eq:D_local_coordinates}
\end{equation}
In these coordinates, the super-ASW family is flattened to
$\eta_2=0$, i.e., $d_1(0,\eta_1,0)\equiv0$,
while the off-$\Gamma$ super-BIC family remains a curved
branch emanating from {\sf D}, as illustrated in
Fig.~\ref{superaswDgeometry}(c).

At {\sf D}, $d_j=0$ for $j<5$ but $d_5\neq0$.
Keeping the leading terms near {\sf D} gives
\begin{equation}
	d
	\sim
	\delta
	\left[
	d_5\big|_{\sf D}\delta^4
	+
	(\partial_{\eta_1}d_3)\big|_{\sf D}\eta_1\delta^2
	+
	(\partial_{\eta_2}d_1)\big|_{\sf D}\eta_2
	\right],
	\label{eq:D_radiation_local}
\end{equation}
Therefore, for $\delta\neq0$, the leading
condition for a nearby off-$\Gamma$ BIC can be written as
\begin{equation}
	\delta^4-a\eta_1\delta^2+b\eta_2=0,
	\label{eq:D_normal_form}
\end{equation}
where
\begin{equation}
	a=-\frac{(\partial_{\eta_1}d_3)}{d_5},
	\quad
	b=\frac{(\partial_{\eta_2}d_1)}{d_5}.
	\label{eq:D_ab}
\end{equation}
For $a\neq0$ and $b\neq0$, Eq.~\eqref{eq:D_normal_form}
describes a cusp-like geometry, with its orientation determined
by the signs of $a$ and $b$.
For {\sf D}, direct numerical evaluation gives $a>0$ and $b>0$.

An intuitive feature of this cusp-like geometry is
that the super-ASW ($\eta_2=0$) and off-$\Gamma$
super-BIC families divide the structural parameter
space near {\sf D} into three regions, I, II, and III,
with different numbers of off-$\Gamma$ BICs for
fixed structural parameters, as shown in
Figs.~\ref{superaswDgeometry}(a)--(c).
The numbers of nearby off-$\Gamma$ BICs and
the local boundaries of these regions can be
determined to leading order from the real roots
of Eq.~\eqref{eq:D_normal_form}.
For $a>0$ and $b>0$, this equation has two
nonzero real roots when $\eta_2<0$,
four when $\eta_1>0$ and
$0<\eta_2<a^2\eta_1^2/(4b)$,
and none in the remaining region with $\eta_2>0$.
These conditions provide leading-order local
descriptions of regions I, II, and III, respectively.
The boundary between regions II and III is the
off-$\Gamma$ super-BIC family, characterized by
the exact conditions $d=0$ and
$\partial_\delta d=0$ at $\delta\neq0$.
Applying these conditions to the leading-order
equation~\eqref{eq:D_normal_form} gives
\begin{equation}
	\delta^2
	\sim
	\frac{a}{2}\eta_1,
	\quad
	\eta_2
	\sim
	\frac{a^2}{4b}\eta_1^2,
	\quad
	\eta_1>0.
	\label{eq:D_offgamma_family}
\end{equation}
Equation~\eqref{eq:D_offgamma_family} shows
that the off-$\Gamma$ super-BIC family is
quadratically tangent to the super-ASW family
$\eta_2=0$ at {\sf D}.

The cusp geometry gives rise to different bifurcation
behaviors along three generic structural
paths, each passing through a different super-ASW.
We consider paths with fixed $\eta_1<0$, $\eta_1=0$,
and $\eta_1>0$, as illustrated in
Figs.~\ref{superaswDgeometry}(d)--(f).
For $\eta_1<0$, increasing $\eta_2$ takes the section
directly from region I to region III, reducing the number
of off-$\Gamma$ BICs from two to zero.
For $\eta_1=0$, the path passes through {\sf D}, where
Eq.~\eqref{eq:D_normal_form} reduces to
$\delta^4+b\eta_2=0$.
Thus, $\eta_2\sim\delta^4$ in region I,
consistent with Fig.~\ref{bifurcation}(f).
For $\eta_1>0$, the section passes successively through
regions I, II, and III, and the number of off-$\Gamma$
BICs changes from two to four and then to zero.
The combined pitchfork and saddle-node bifurcations
produce an ``M''-shaped pattern of BIC branches
in the $(\beta,\eta_2)$ plane.

We next examine two non-generic structural paths
passing through {\sf D}.
The first path follows the off-$\Gamma$
super-BIC family.
Along this path, two symmetry-related
off-$\Gamma$ super-BICs approach $\delta=0$
and merge with the super-ASW at {\sf D},
as shown in Figs.~\ref{3Doffgammasuperbic}
and \ref{superaswDgeometry}(a)--(c).
The second path follows the super-ASW family itself,
$\eta_2=0$.
For the additional off-$\Gamma$ BICs,
Eq.~\eqref{eq:D_normal_form} gives
$\delta^2\sim a\eta_1$.
Since $a>0$, no nearby off-$\Gamma$ BIC exists
for $\eta_1<0$, whereas one pair exists for
$\eta_1>0$.
As $\eta_1\to0^+$, this pair approaches
$\delta=0$ and merges with the super-ASW
at {\sf D}, as shown in
Fig.~\ref{superaswDgeometry}(g).
Thus, a bifurcation occurs along the super-ASW
family itself, where the leading radiation
of nearby resonant states changes from third
order to fifth order at {\sf D}.

Our results not only reveal the local BIC geometry
near {\sf D}, but also explain why realizing such
a super-ASW through BIC merging is challenging.
A conventional approach is to merge four
off-$\Gamma$ BICs with the symmetry-protected
BIC at $\delta=0$~\cite{HLuo2023PRA,LLiu2024OE}.
This requires approaching {\sf D} through region II,
whose width in the $\eta_2$ direction decreases
quadratically as $\eta_1\to0^+$, according to
Eq.~\eqref{eq:D_offgamma_family}.
The structural path must therefore be tangent
to the super-ASW family at {\sf D}.
The two non-generic paths discussed above provide
alternative merging scenarios, but both must be
carefully chosen.
Even for a generic structural path, reaching {\sf D}
requires passing through an isolated point
in the 2D structural parameter space, which is
itself nontrivial.

\section{Super-BICs in 3D biperiodic structures}

Finally, we extend the preceding results to vectorial
fields in lossless and reciprocal 3D biperiodic structures.
The essential difference is that the leading radiation depends
on the direction of the wavevector detuning.
Nevertheless, the theoretical framework developed
in the preceding sections remains applicable.
Rather than repeating the detailed derivations,
we present the main results.
In subsections A--C, we classify super-BICs
and determine their codimensions in structural
parameter space using the theory of parametric
dependence.
In subsection D, we present numerical examples
that verify the classification and codimension results
and demonstrate the bifurcation behaviors predicted
by the vectorial extension of our theory.
The super-BICs in these examples are located using
the direct computational method developed in Sec.~V.

We continue to assume up-down mirror symmetry
and that only the zeroth diffraction order is
propagating in each half-space.
Let $({\bm\kappa}_*,\omega_*,{\bm\Psi}_*)$
denote a nondegenerate BIC and parameterize
a nearby Bloch wavevector by
${\bm\kappa}={\bm\kappa}_*+\delta{\bm\theta}/L$,
where ${\bm\theta}=(\cos\theta,\sin\theta)$
and $-\pi\leq\theta<\pi$.
Based on the perturbation theory in Sec.~\Rmnum{2}
and Ref.~\cite{NZhang2026arXiv}, the frequency
and field corrections can be analyzed order by order,
allowing us to identify the leading radiative
field correction.
We construct the propagating diffraction channels
at (${\bm\kappa}_*$, $\omega_*$).
Projecting the far fields of the nearby resonant
state and the leading radiative field correction
onto these fixed channels defines the radiation
vectors ${\bm d}({\bm\kappa})$ and
${\bm d}_\mu(\theta)$, respectively.
The leading radiation satisfies
\begin{equation}
	{\bm d}({\bm\kappa})
	\sim
	\delta^\mu{\bm d}_\mu(\theta).
	\label{eq:3d_leading_radiation}
\end{equation}

The angular structure and topological properties
of ${\bm d}_\mu(\theta)$ have been studied in detail
in Ref.~\cite{NZhang2026Topology}.
Here, we focus on ultrahigh-$Q$ resonances near
super-BICs.
We define
${\cal Z}_\mu=\{\theta:
{\bm d}_\mu(\theta)={\bm 0}\}$.
For $\theta\notin{\cal Z}_\mu$,
$Q\sim Q_{2\mu}(\theta)/\delta^{2\mu}$,
whereas along $\theta\in{\cal Z}_\mu$
the asymptotic behavior is determined by
higher-order radiation.
We use the relevant results to classify
super-BICs with $\mu\geq2$ and super-BICs with
$\mu=1$ and ${\cal Z}_1\neq\varnothing$,
and determine their corresponding codimensions
without repeating the derivations.

For an at-$\Gamma$ BIC, we choose its phase such that
${\cal T}{\bm\Psi}_*={\bm\Psi}_*$,
where ${\cal T}$ is the time-reversal operator.
The leading radiation then satisfies
\begin{equation}
	{\bm d}_\mu(\theta)
	=
	(-1)^{\mu+1}
	{\bm S}\,
	\overline{{\bm d}_\mu}(\theta).
	\label{eq:3d_Gamma_constraint}
\end{equation}
Here ${\bm S}$ is the $2\times2$ reduced
scattering matrix at $(\omega_*,{\bm\kappa}_*)$.
For an off-$\Gamma$ BIC in a structure with
$C_2$ symmetry, we instead choose the phase
such that $C_2{\cal T}{\bm\Psi}_*={\bm\Psi}_*$,
which gives
\begin{equation}
	{\bm d}_\mu(\theta)
	=
	-{\bm S}\,
	\overline{{\bm d}_\mu}(\theta).
	\label{eq:3d_offGamma_constraint}
\end{equation}

\subsection{$C_\nu$ symmetry: classification and codimension}

We first consider nondegenerate at-$\Gamma$ BICs with $C_\nu$
symmetry, where $\nu=3,4,$ or $6$, and let $\tau=\pm1$ denote the
rotational character of the BIC. The reduced scattering matrix
satisfies ${\bm S}=s{\bm I}$ with $|s|=1$.
Together with Eq.~\eqref{eq:3d_Gamma_constraint}, this implies that
${\bm d}_\mu(\theta)$ can be made real for all $\theta$ by a fixed
phase transformation. The leading radiation can then be written as
\begin{equation}
	{\bm d}_\mu(\theta)
	=
	\sum_{m\in{\cal M}^{\nu,\tau}_\mu}
	{\bm R}_{m\theta}{\bm c}_{\mu,m},
	\label{eq:3d_angular_decomposition}
\end{equation}
where ${\bm c}_{\mu,m}$ are real constant vectors, and
${\cal M}^{\nu,\tau}_\mu$ for the lowest few orders are listed in
Table~\ref{tab:3d_harmonics}.
A detailed derivation of
Eq.~\eqref{eq:3d_angular_decomposition} can be found in
Ref.~\cite{NZhang2026Topology}.
\begin{table}[t]
	\centering
	\caption{
		Allowed angular harmonics ${\cal M}^{\nu,\tau}_\mu$ for
		nondegenerate at-$\Gamma$ BICs with $C_\nu$ symmetry.
	}
	\label{tab:3d_harmonics}
	\begin{tabular}{c|c|c|c|c|c}
		\hline\hline
		& $C_3$ & \multicolumn{2}{c|}{$C_4$}
		& \multicolumn{2}{c}{$C_6$}\\
		$\mu$
		& $\tau=1$
		& $\tau=1$ & $\tau=-1$
		& $\tau=1$ & $\tau=-1$\\
		\hline
		1 & $1$       & $1$         & $-1$          & $1$        & --\\
		2 & $-2$      & --          & --            & --         & $-2$\\
		3 & $1$       & $-3,1$      & $-1,3$        & $1$        & --\\
		4 & $-2,4$    & --          & --            & --         & $-2,4$\\
		5 & $-5,1$    & $-3,1,5$    & $-5,-1,3$     & $-5,1$     & --\\
		6 & $-2,4$    & --          & --            & --         & $-2,4$\\
		\hline\hline
	\end{tabular}
\end{table}

Since ${\bm d}_\mu(\theta)$ is real, we introduce
\begin{equation}
	f_\mu(\theta)
	=
	d_{\mu,x}(\theta)
	+
	i d_{\mu,y}(\theta)
	=
	\sum_{m\in{\cal M}^{\nu,\tau}_\mu}
	a_{\mu,m}e^{im\theta}.
	\label{eq:3d_fmu}
\end{equation}
The angular zero set is therefore
\begin{equation}
	{\cal Z}_\mu
	=
	\left\{
	\theta:
	f_\mu(\theta)=0
	\right\}.
	\label{eq:3d_zero_set_Cn}
\end{equation}
If the leading radiation contains only one harmonic,
${\cal Z}_\mu=\varnothing$, and
$Q\sim1/\delta^{2\mu}$ in every direction.
When multiple harmonics are present, the angular
zeros can be determined directly from their coefficients.
In particular, if two harmonics $m_-<m_+$ are present,
angular zeros occur when their nonzero coefficients satisfy
$|a_{\mu,m_-}|=|a_{\mu,m_+}|$.
In this case,
\begin{equation}
	{\cal Z}_\mu
	=
	\left\{
	\theta\in[-\pi,\pi):
	e^{i(m_+-m_-)\theta}
	=
	-\frac{a_{\mu,m_-}}{a_{\mu,m_+}}
	\right\},
	\label{eq:3d_two_harmonic_zero_set}
\end{equation}
giving $m_+-m_-$ equally spaced directions.
Away from these directions,
$Q\sim1/\delta^{2\mu}$, whereas along them
higher-order radiation determines the asymptotic scaling.

We illustrate the classification of super-BICs
for the first few allowed values of $\mu$.
The classification for higher values of $\mu$
follows similarly.
For $C_6$ symmetry with $\tau=-1$, first-order
radiation is forbidden, and the second-order
radiation contains only the harmonic $m=-2$.
Thus, every at-$\Gamma$ BIC in this class
is automatically a super-BIC.
When the second-order coefficient is nonzero,
$Q\sim1/\delta^4$ in all directions,
without angular zeros.
If the second-order radiation vanishes identically,
the next allowed order is $\mu=4$, with
harmonics $m=-2,4$, allowing the presence of
angular zeros.

For $C_6$ symmetry with $\tau=1$, first-order
radiation is allowed and contains only $m=1$,
giving the generic scaling $Q\sim1/\delta^2$.
If this radiation vanishes identically, the next
allowed order is $\mu=3$, which also contains
only $m=1$.
The resulting super-BIC therefore has
$Q\sim1/\delta^6$ in every direction
without angular zeros.
If the third-order radiation also vanishes,
the leading radiation generally occurs at $\mu=5$,
with harmonics $m=-5,1$, allowing the presence
of angular zeros.

For $C_4$ symmetry, both $\tau=1$ and $\tau=-1$
allow first-order radiation with a single harmonic.
Thus, a generic at-$\Gamma$ BIC has
$Q\sim1/\delta^2$ in every direction.
If the first-order radiation vanishes identically,
the second-order radiation is forbidden by symmetry,
and the leading radiation generally occurs at $\mu=3$.
For $\tau=1$, the allowed harmonics are $m=-3,1$,
while for $\tau=-1$, they are $m=-1,3$.
In both cases, angular zeros are possible.

For $C_3$ symmetry, the first super-BIC order is
$\mu=2$, with a single harmonic and
$Q\sim1/\delta^4$ in every direction.
Angular zeros first become possible at $\mu=4$,
where two harmonics are allowed.

We next determine the codimension $n$ of each class. 
The rigorous
analysis developed in Sec.~\Rmnum{4} can be extended
directly to the 3D setting. Here we state the resulting counting
rules without repeating the full derivation. For the radiation of
nearby resonant states to begin at order $\mu$, all symmetry-allowed
radiation coefficients at lower orders vanish. 
We consider $C_{\nu v}$ structures and choose
the coordinate system such that one of the in-plane mirror symmetries
is $y\to-y$. This symmetry reduces each allowed harmonic to one
independent real coefficient. The codimension is therefore
\begin{equation}
	n_{\nu,\tau}(\mu)
	=
	\sum_{j=1}^{\mu-1}
	\left|{\cal M}^{\nu,\tau}_j\right|.
	\label{eq:3d_codimension}
\end{equation}
The resulting values for the first few $\mu$
are listed in Table~\ref{tab:3d_codimension}. 
If one further requires
${\cal Z}_\mu\neq\varnothing$, the angular-zero condition generically
introduces one additional real constraint. Thus, a super-BIC of
codimension $n_{\nu,\tau}(\mu)$ with an angular zero generically has
codimension $n_{\nu,\tau}(\mu)+1$.
Without in-plane mirror symmetry, each allowed harmonic
generally has two independent real coefficients.
Consequently, the codimension is twice that of
the corresponding mirror-symmetric case.
\begin{table}[t]
	\centering
	\caption{
		Codimension $n_{\nu,\tau}(\mu)$ of nondegenerate at-$\Gamma$
		super-BICs with $C_{\nu v}$ symmetry for the lowest few
		values of $\mu$.
	}
	\label{tab:3d_codimension}
	\begin{tabular}{c|c|c|c|c|c}
		\hline\hline
		& $C_{3v}$ & \multicolumn{2}{c|}{$C_{4v}$}
		& \multicolumn{2}{c}{$C_{6v}$}\\
		$\mu$
		& $\tau=1$
		& $\tau=1$ & $\tau=-1$
		& $\tau=1$ & $\tau=-1$\\
		\hline
		1 & $0$ & $0$ & $0$ & $0$ & --\\
		2 & $1$ & --  & --  & --  & $0$\\
		3 & $2$ & $1$ & $1$ & $1$ & --\\
		4 & $3$ & --  & --  & --  & $1$\\
		5 & $5$ & $3$ & $3$ & $2$ & --\\
		6 & $7$ & --  & --  & --  & $3$\\
		\hline\hline
	\end{tabular}
\end{table}

\subsection{At-$\Gamma$ BICs with $C_2$ symmetry}
We next consider structures with $C_2$ rotational
symmetry.
We note that the 2D structures considered in the
preceding sections also possess $C_{2v}$ symmetry.
Our restriction to scalar fields in those sections
was intended to facilitate the presentation of
the theoretical framework and the underlying
physical picture.
In the more general vectorial setting, our analysis
applies to arbitrary $\mu$, but in this subsection we focus on two cases:
super-BICs with $\mu\ge2$ and super-BICs with
$\mu=1$ and ${\cal Z}_1\ne\varnothing$.
For $\tau=-1$, first-order radiation is forbidden
by symmetry, so every at-$\Gamma$ BIC in this class
is automatically a super-BIC with $\mu\ge2$.
Such super-BICs have codimension two, which is
reduced to one if an additional in-plane mirror
symmetry is imposed.
We therefore focus on at-$\Gamma$ BICs with $\tau=1$
and determine the codimensions associated with
the two cases.

For the lossless and reciprocal system, ${\bm S}$ is symmetric and
unitary. 
Its Takagi factorization can therefore be written as ${\bm S}={\bm U}{\bm U}^{\sf T}$,
where ${\bm U}$ is unitary. For $\mu=1$,
Eq.~\eqref{eq:3d_Gamma_constraint} then implies that
${\bm U}^{\dagger}{\bm d}_1(\theta)$ is real. Since the first-order
radiation is linear in $\cos\theta$ and $\sin\theta$, it can be
written as
\begin{equation}
	{\bm U}^{\dagger}{\bm d}_1(\theta)
	=
	{\bm V}
	\begin{bmatrix}
		\cos\theta\\
		\sin\theta
	\end{bmatrix},
	\quad
	{\bm V}\in\mathbb{R}^{2\times2}.
	\label{eq:C2_first_order}
\end{equation}
For a generic full-rank ${\bm V}$,
${\cal Z}_1=\varnothing$.
When ${\bm V}$ has rank one, its null vector
determines an angular-zero direction $\theta_c$.
Hence ${\cal Z}_1=\{\theta_c,\theta_c+\pi\}$,
where the two angles are understood in $[-\pi,\pi)$. 
Thus, ${\cal Z}_1\neq\varnothing$ imposes one real condition, 
while
the leading radiation order remains $\mu=1$ for all other directions.

For the first-order radiation to vanish in all directions, one
requires ${\bm V}={\bm0}$.
With only $C_2$ symmetry, ${\bm V}$ contains four independent real
coefficients, so this condition has codimension four. Since only odd
radiation orders are allowed for $\tau=1$, the next possible leading
order is $\mu=3$.

If an additional in-plane mirror symmetry is imposed, we choose the
coordinates such that the mirror transformation is $y\to-y$.
Depending on the mirror character of the BIC, ${\bm V}$ takes one of
the forms
\begin{equation}
	{\bm V}
	=
	\begin{bmatrix}
		a&0\\
		0&b
	\end{bmatrix}
	\quad\text{or}\quad
	{\bm V}
	=
	\begin{bmatrix}
		0&a\\
		b&0
	\end{bmatrix},
	\label{eq:C2_mirror_A}
\end{equation}
where $a$ and $b$ are real. An angular zero occurs when either
$a=0$ or $b=0$, and the corresponding zero set is
\begin{equation}
	{\cal Z}_1
	=
	\{0,-\pi\}
	\quad\text{or}\quad
	\left\{-\frac{\pi}{2},\frac{\pi}{2}\right\}.
	\label{eq:C2_mirror_zero_set}
\end{equation}
Thus, a first-order angular zero remains a codimension-one condition,
whereas eliminating the first-order radiation in all directions
requires $a=b=0$ and has codimension two.

\subsection{Degenerate and off-$\Gamma$ BICs}

The framework can also be extended to degenerate and off-$\Gamma$
BICs. In particular, the degenerate perturbation theory developed in
Sec.~\Rmnum{2}B applies directly to the 3D setting.

We first consider doubly degenerate at-$\Gamma$ BICs in
$C_{4v}$ and $C_{6v}$ structures.
BICs belonging to the $E$ representation of $C_{4v}$ or
the $E_1$ representation of $C_{6v}$
are typically super-BICs with $\mu=2$ and can be realized
by tuning one structural parameter,
analogous to the degenerate SSW case discussed in
Sec.~\Rmnum{2}B~\cite{AKostyukov2022PRB,TOchiai2024PRA}.
In contrast, BICs belonging to the $E_2$ representation of $C_{6v}$ 
generally have nonzero first-order radiation, 
giving $\mu=1$ for the nearby resonant bands, 
analogous to the degenerate ASW case.
Doubly degenerate BICs belonging to the $E$ representation of $C_{4v}$ have also been employed in BIC lasers~\cite{AKodigala2017Nature}.

For off-$\Gamma$ BICs at high-symmetry points or on
high-symmetry lines, the classification and codimension
can be determined using the corresponding little-group
symmetries.
Away from these locations, both can generally be
determined directly from the radiation coefficients.

\subsection{Numerical examples}

We numerically verify the codimension results
and demonstrate the bifurcation behaviors of
super-BICs with $\mu=3$ in $C_{4v}$-symmetric structures.
We consider at-$\Gamma$ BICs in the $A_2$ representation,
with rotational character $\tau=1$.
According to Tables~\ref{tab:3d_harmonics} and
\ref{tab:3d_codimension}, this class of super-BICs
has codimension one.
Requiring an angular zero of the third-order radiation
introduces one additional constraint.
Other classes of super-BICs can be investigated
in a similar manner.

We consider the square lattice of dielectric rods shown in
Fig.~\ref{structure}(b).
The rods have refractive index $3.5$ and are suspended in air.
Since $n=1$, these super-BICs form curves in the 2D structural
parameter space $(r,h)$, where $r$ and $h$ denote
the rod radius and height, respectively.
Using the direct method of Sec.~\Rmnum{5}, we trace
the super-BIC family by scanning $r$ and solving for
$h$ and $\omega$.

For $\mu=1$, the only allowed angular harmonic is $m=1$,
so suppressing the first-order radiation is equivalent to
$a_{1,1}=0$.
We therefore solve
\begin{equation}
	{\bm G}(\omega,h;r)
	=
	\begin{bmatrix}
		\lambda_0\\
		a_{1,1}
	\end{bmatrix}
	=
	{\bm 0}.
	\label{eq:3d_direct_system}
\end{equation}
For $\mu=3$, the allowed angular harmonics are $m=-3$ and
$m=1$, giving
\begin{equation}
	f_3(\theta)
	=
	a_{3,-3}e^{-i3\theta}
	+
	a_{3,1}e^{i\theta}.
	\label{eq:C4_mu3_f3}
\end{equation}
Super-BICs with ${\cal Z}_3\ne\varnothing$
can be computed by solving
\begin{equation}
	{\bm G}(\omega,h,r)
	=
	\begin{bmatrix}
		\lambda_0\\
		a_{1,1}\\
		|a_{3,-3}|-|a_{3,1}|
	\end{bmatrix}
	=
	{\bm 0}
	\label{eq:3d_transition_system}
\end{equation}
for $\omega$, $h$, and $r$.

\begin{figure*}[htbp]
	\centering
	\includegraphics[scale=0.33]{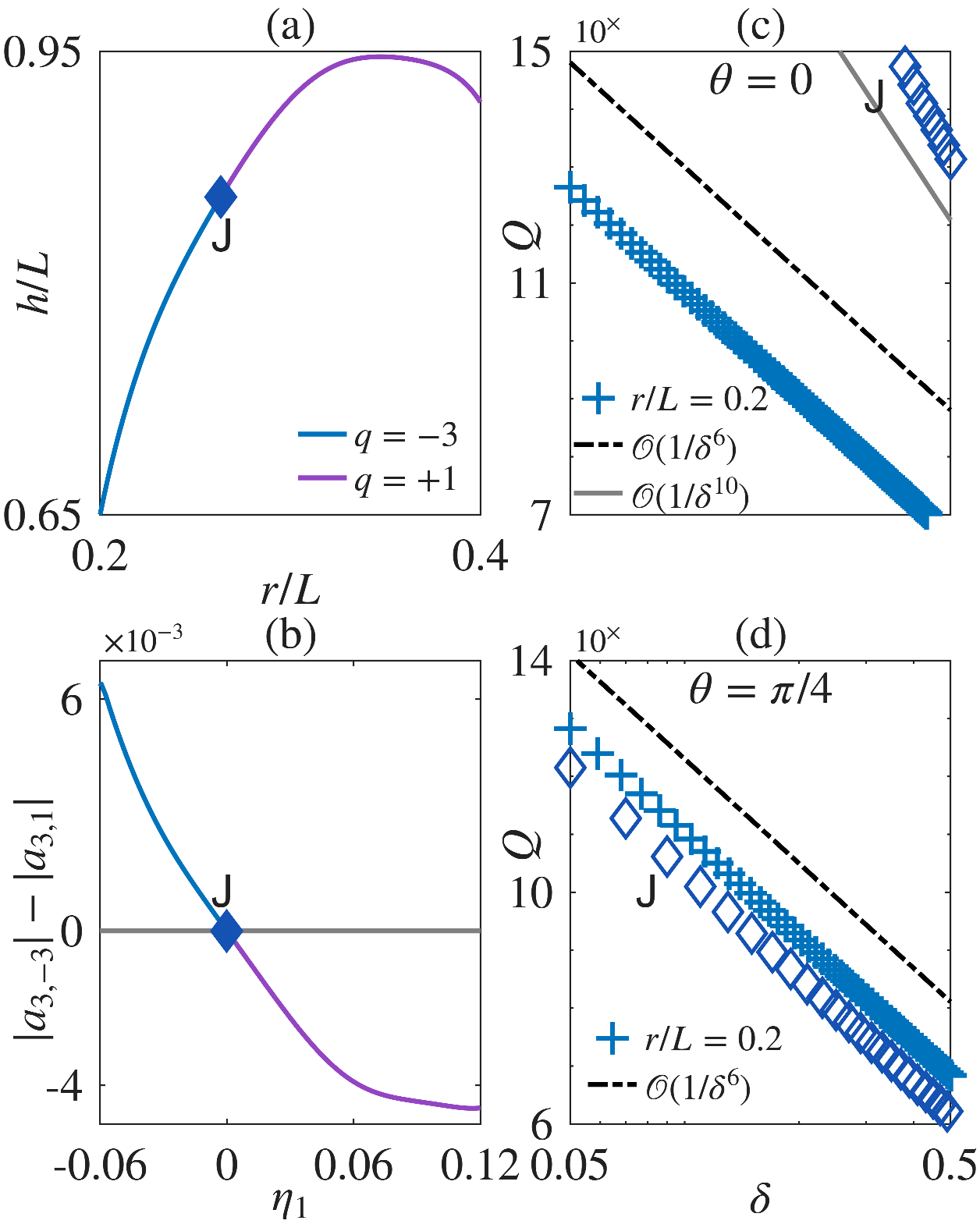}
	\includegraphics[scale=0.33]{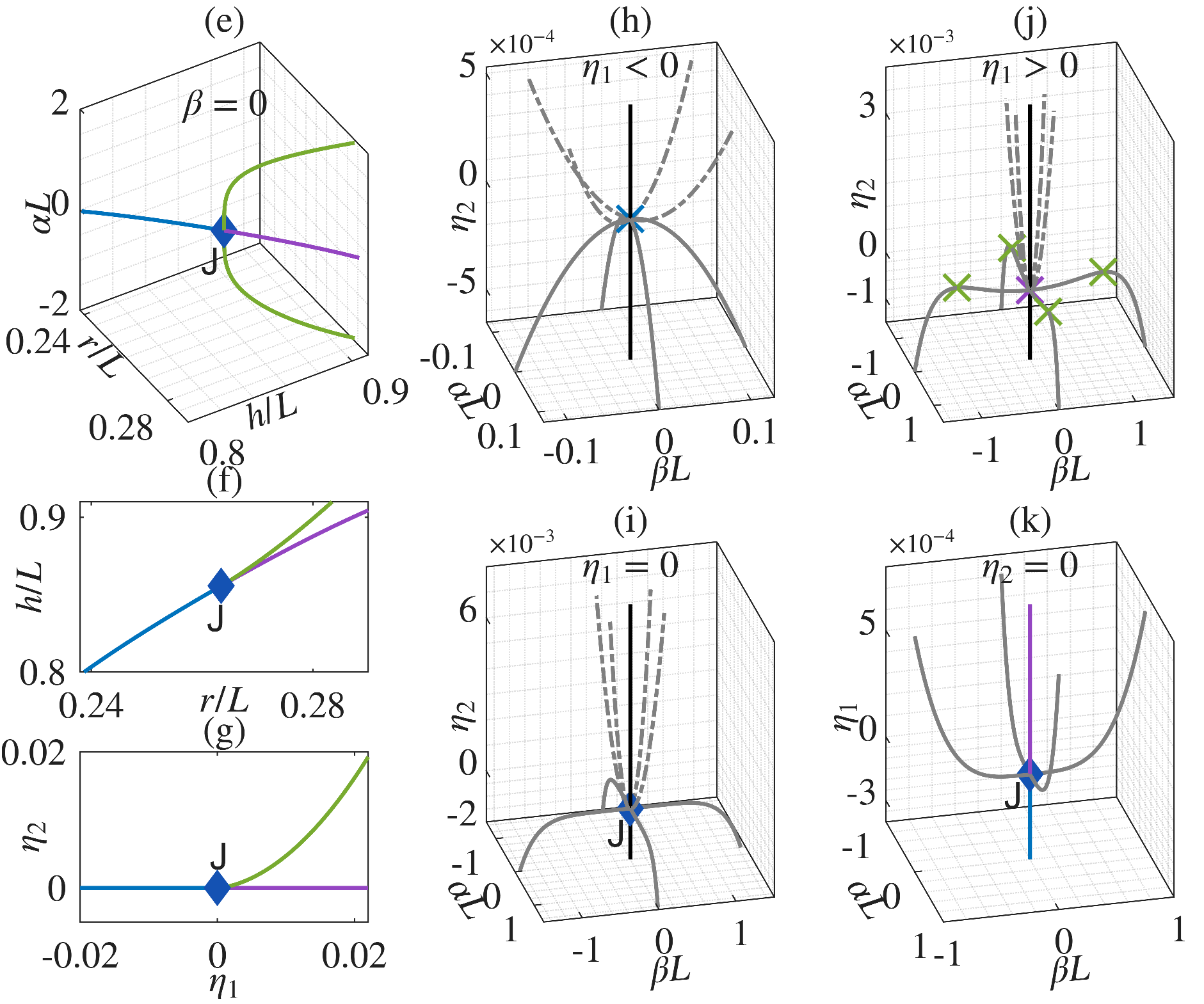}
	\caption{
		Numerical results for $\mu=3$ super-BICs in the $C_{4v}$ structure shown in Fig.~\ref{structure}(b).
		(a) Super-BIC family in the $(r,h)$ parameter space.
		The blue and purple segments correspond to $q=-3$ and $q=+1$, respectively, and the blue diamond marks the transition point $\sf J$.
		(b) Difference $|a_{3,-3}|-|a_{3,1}|$ along the family, parameterized by the local coordinate $\eta_1$.
		The angular-zero condition is satisfied at $\sf J$, where $\eta_1=0$.
		(c),(d) $Q$ factors of nearby resonant states along $\theta=0$ and $\theta=\pi/4$, respectively.
		The results at $r/L=0.2$ confirm the generic $Q\sim1/\delta^6$ scaling, while at $\sf J$ the scaling along $\theta=0$ becomes $Q\sim1/\delta^{10}$.
		(e) At-$\Gamma$ and off-$\Gamma$ super-BIC families in $(r,h,\alpha)$ with $\beta=0$, where the off-$\Gamma$ families are shown in green.
		(f),(g) Super-BIC families in $(r,h)$ and $(\eta_1,\eta_2)$, respectively.
		(h)--(j) BIC bifurcations for fixed $\eta_1<0$, $\eta_1=0$, and $\eta_1>0$, respectively.
		(k) BIC bifurcation for $\eta_2=0$.
	}
	\label{superasw3Dgeometry}
\end{figure*}
The computed super-BIC family is shown in
Fig.~\ref{superasw3Dgeometry}(a).
The winding number of the leading radiation is $q=-3$
when $|a_{3,-3}|>|a_{3,1}|$ and $q=+1$ when
$|a_{3,-3}|<|a_{3,1}|$~\cite{NZhang2026Topology}.
The transition point ${\sf J}$, characterized by
$|a_{3,-3}|=|a_{3,1}|$, is located at
$(r_J,h_J)=(0.2634,0.8557)L$, with $\omega L=2.927$.
Let $h=h_*(r)$ denote the super-BIC family, with
$h_*(r_J)=h_J$.
We introduce the local coordinates
\begin{equation}
	\eta_1=\frac{r-r_J}{L},
	\quad
	\eta_2=\frac{h-h_*(r)}{L}.
	\label{eq:J_local_coordinates}
\end{equation}
The super-BIC family corresponds to $\eta_2=0$,
and ${\sf J}$ is located at $\eta_1=\eta_2=0$.
Figure~\ref{superasw3Dgeometry}(b) shows
$|a_{3,-3}|-|a_{3,1}|$ as a function of $\eta_1$,
which vanishes at ${\sf J}$.
At this point, $a_{3,-3}=-a_{3,1}$, so
\begin{equation}
	{\cal Z}_3
	=
	\left\{
	-\pi+\frac{j\pi}{2},\quad j=0,1,2,3
	\right\}.
\end{equation}

At ${\sf J}$, the third-order radiation vanishes along
all four directions in ${\cal Z}_3$, and further calculation
shows that the leading radiation appears at fifth order,
giving $Q\sim1/\delta^{10}$.
Along other directions, the leading radiation remains
third order, with $Q\sim1/\delta^6$.
In contrast, other super-BICs on the $\mu=3$ family
exhibit $Q\sim1/\delta^6$ in all directions.
These results are confirmed in
Figs.~\ref{superasw3Dgeometry}(c) and (d), which compare
${\sf J}$ with a super-BIC at $r/L=0.2$
along $\theta=0$ and $\theta=\pi/4$, respectively.

We note that the super-BIC ${\sf J}$ is similar to
the $\mu=5$ super-ASW ${\sf D}$ discussed in
Secs.~\Rmnum{5} and \Rmnum{6} and illustrated in
Figs.~\ref{3Doffgammasuperbic} and
\ref{superaswDgeometry}.
Both are special points along families of super-BICs
where the leading radiation order increases from
three to five along specific directions.
In particular, the off-$\Gamma$ super-BICs located
along these directions merge with the at-$\Gamma$
super-BIC.
Consequently, when restricted to $(r,h,\alpha)$ with
$\beta=0$, where ${\bm\kappa}=(\alpha,\beta)$,
the local BIC set near ${\sf J}$ exhibits the same cusp
geometry as that near ${\sf D}$ shown in
Fig.~\ref{superaswDgeometry}(a).

Rather than plotting the full local BIC set shown in
Fig.~\ref{superaswDgeometry}(a), we focus on the at-$\Gamma$
and off-$\Gamma$ super-BIC families, following
Fig.~\ref{3Doffgammasuperbic}(a) and
Figs.~\ref{superaswDgeometry}(b) and (c).
Figure~\ref{superasw3Dgeometry}(e) shows these families
in $(r,h,\alpha)$ with $\beta=0$, while
Figs.~\ref{superasw3Dgeometry}(f) and (g) show their
projections onto $(r,h)$ and $(\eta_1,\eta_2)$, respectively.
The two families meet tangentially at ${\sf J}$,
reproducing the geometry near ${\sf D}$ and dividing the
neighboring structural parameter space into three regions
with different numbers of off-$\Gamma$ BICs in the
$\beta=0$ section.

Similar to the case of ${\sf D}$, the corresponding
bifurcations along structural paths with fixed
$\eta_1<0$, $\eta_1=0$, $\eta_1>0$, and $\eta_2=0$
are illustrated in Figs.~\ref{superasw3Dgeometry}(h)--(k),
respectively.
The off-$\Gamma$ BIC branches can be divided into
two families associated with $\theta=0$ and
$\theta=\pi/4$, together with their symmetry-related
counterparts under $C_{4v}$.
The $\theta=0$ family exhibits bifurcation behaviors
similar to those near ${\sf D}$, including a
higher-order pitchfork bifurcation at ${\sf J}$
and an ``M''-shaped pattern for fixed $\eta_1>0$.
In contrast, the $\theta=\pi/4$ family exhibits no
corresponding critical change at ${\sf J}$, since
the third-order radiation remains nonzero
along these directions.

\section{Conclusion}

The higher-order $Q$-factor scaling associated with super-BICs
can be traced back to 2017, when resonances near certain special
BICs were shown to exhibit $Q\sim1/\delta^4$, in contrast to
the usual $Q\sim1/\delta^2$ behavior~\cite{LYuan2017PRA}.
BIC merging had already been considered in the topological
description of BICs~\cite{BZhen2014PRL}.
In 2017, its connection to enhanced $Q$-factor scaling was
made more explicit, with resonances near a BIC annihilation
point shown to exhibit $Q\sim1/\delta^4$~\cite{EBulgakov2017PRA}.
An important experimental advance came in 2019, when
the merging of multiple BICs was demonstrated to produce
ultrahigh-$Q$ resonances with
$Q\sim1/\delta^6$~\cite{JJin2019Nature}.
However, realizing higher-order super-BICs through merging
generally requires tuning multiple structural parameters
while tracking an increasing number of BICs,
making this approach increasingly challenging.

Another approach characterizes super-BICs through
the vanishing of radiation coefficients in the perturbation
expansion of nearby resonant states~\cite{LYuan2017PRA,NZhang2025PRL}.
These coefficients can be determined directly from the BIC.
More recently, an FMM-based formulation has enabled
the computation of the required field corrections and
radiation coefficients in layered periodic
structures~\cite{NZhang2026arXiv}.
Together, these developments provide a theoretical and
computational basis for the systematic study of super-BICs.

In this work, we establish a systematic theoretical and
computational framework for studying super-BICs.
We classify super-BICs based on $\mu$ and symmetry,
including in-plane rotational and reflection symmetries.
The unfolding analysis characterizes the bifurcations
of nearby generic BICs under structural variations
and determines the minimum number $n$ of additional
tunable structural parameters required to recover
super-BICs of the same class.
These results predict how different classes of super-BICs
are distributed in structural parameter space.
Our direct computational method allows super-BICs
to be located without tracking BIC-merging processes.
Numerical examples demonstrate that higher-order
super-BICs with $\mu=5$ can be located efficiently
and that their families can be traced in
higher-dimensional parameter spaces.

The local geometry of the BIC set in
${\cal P}\times\mathrm{BZ}$ provides a unified
interpretation of the unfolding of super-BICs.
The number $n$ is the codimension of the corresponding
super-BIC submanifold within the BIC manifold,
while structural paths select sections of the BIC set
that determine the bifurcations of nearby generic BICs.
We identify saddle points, elliptic points, and cusp
geometries near super-BICs.
Different structural paths through the same super-BIC
may produce qualitatively different bifurcation behaviors,
including linear bifurcations or the absence of bifurcation.
This geometric viewpoint provides a new understanding
of BIC merging and splitting phenomena and explains
why realizing higher-order super-BICs through merging
can require carefully chosen structural paths.

Overall, our results provide a unified understanding
of super-BICs and a theoretical and computational basis
for realizing ultrahigh-$Q$ resonances,
with potential applications in enhancing
light-matter interactions.

\bibliography{prb_superbic}
\end{document}